\documentclass[11pt,a4paper]{article}
\pdfoutput=1
\usepackage{jheppub}
\usepackage{slashed}
\usepackage{comment}
\usepackage{makecell}

\usepackage[english]{babel}

\usepackage{amsmath,bm}
\usepackage{amssymb}
\usepackage{graphicx}
\usepackage{ulem}
\usepackage{multirow}
\usepackage{xcolor,color}
\usepackage{booktabs}
\allowdisplaybreaks

\usepackage{siunitx}
\usepackage{url}

\definecolor{gesfblack}{rgb}{0,0,0}

\definecolor{gesfblue}{rgb}{0.08,0.42,0.76}

\definecolor{gesfgreen}{rgb}{0.09, 0.45, 0.27}

\definecolor{gesfgrey}{rgb}{0.5,0.5,0.5}

\definecolor{gesflanse}{rgb}{0.00,0.50,0.50}

\definecolor{gesfpurple}{rgb}{0.47,0.19,0.42}

\definecolor{gesfred}{rgb}{1,0,0}

\definecolor{gesfwhite}{rgb}{1,1,1}

\definecolor{gesfyellow}{rgb}{0.7,0.4,0.3}

\newcommand{\gsec}[1]{{\hypersetup{linkcolor=red}Sec.\,\ref{#1}\hypersetup{linkcolor=blue}}}

\newcommand{\geqn}[1]{\hypersetup{linkcolor=blue}Eq.\,(\ref{#1})\hypersetup{linkcolor=blue}}
\newcommand{\gfig}[1]{{\hypersetup{linkcolor=violet}Fig.\,\ref{#1}\hypersetup{linkcolor=blue}}}
\newcommand{\gtab}[1]{{\hypersetup{linkcolor=gesflanse}Table~\ref{#1}\hypersetup{linkcolor=blue}}}

\definecolor{Orange}{cmyk}{0,0.61,0.87,0}
\definecolor{JungleGreen}{cmyk}{0.99,0,0.52,0}
\definecolor{OliveGreen}{cmyk}{0.64,0,0.95,0.40}
\definecolor{Brown}{cmyk}{0,0.81,1,0.60}
\definecolor{RoyalBlue}{cmyk}{0.71,0.53,0,0.12}
\definecolor{Gray}{cmyk}{0,0,0,0.40}
\definecolor{LightPink}{cmyk}{0.0,0.25,0,0}
\definecolor{LLightPink}{cmyk}{0.0,0.10,0,0}
\definecolor{LightBlue}{cmyk}{0.25,0,0,0}
\definecolor{LightGray}{cmyk}{0,0,0,0.2}

\graphicspath{{figs/}}

\begin{document}

\title{Probing 5.49\,MeV Solar Axions at Xenon Experiments}

\author{}
\author{Ruofei Feng$^{1}$, Shao-Feng Ge$^{2,3}$, Oleg Titov$^{2,3}$,  Yongchao Zhang$^{1,4}$}

\affiliation{$^1$School of Physics, Southeast University, Nanjing 211189, China}

\affiliation{$^2$State Key Laboratory of Dark Matter Physics, Tsung-Dao Lee Institute \& School of Physics and Astronomy, Shanghai Jiao Tong University, China}
\affiliation{$^3$Key Laboratory for Particle Astrophysics and Cosmology (MOE) \& Shanghai Key Laboratory for Particle Physics and Cosmology, Shanghai Jiao Tong University, Shanghai 200240, China}
\affiliation{$^4$Center for High Energy Physics, Peking University, Beijing 100871, China}

\emailAdd{fengrf@seu.edu.cn, gesf@sjtu.edu.cn, titov$\_$o@sjtu.edu.cn, zhangyongchao@seu.edu.cn}

\abstract{
The monochromatic 5.49\,MeV solar axions induced by the isovector coupling $g_{3aN}$ can be searched for at the dark matter direct detection experiments. In this paper we estimate the prospects of the relevant axion couplings for axion mass $m_{a} <$ 1\,MeV with xenon targets. Given the axion-electron coupling $g_{ae}$, the signal is dominated by the axion-induced $e^+ e^-$ pair production whose cross section is largely enhanced when the axion mass approaches twice of the electron mass. Furthermore, the cross section depends on the atomic number squared $Z^2$. This allows the next-generation xenon experiments to surpass the current Borexino constraints and provide sensitivities competitive with those of the large neutrino detectors such as JUNO and Hyper-Kamiokande.
With an exposure of 200 and 1000 ton$\cdot$yr, the couplings $|g_{3aN} g_{ae}|$
can be probed down to $1.59\times10^{-14}$ and $7.12\times10^{-15}$ at 90\%\,C.L., respectively. If the axion couples to photons,  
the projected sensitivities on $|g_{3aN}g_{a\gamma}|$ can touch down to $6.76\times10^{-12}$\,GeV$^{-1}$ and $3.02\times10^{-12}$\,GeV$^{-1}$, respectively.
}

\maketitle

\section{Introduction}
\label{sec:intro}

The strong CP problem is one of the most profound puzzles in the Standard Model (SM) of particle physics. The quantum chromodynamics (QCD) Lagrangian admits a CP-violating $\bar{\theta}$ parameter, yet experimental measurements constrain its value to be extraordinarily small, $|\bar{\theta}| \lesssim 10^{-10}$, with no natural explanation for this fine-tuning~\cite{Abel:2020pzs}. Peccei and Quinn proposed an elegant solution by introducing a new global $U(1)_\mathrm{PQ}$ symmetry, whose spontaneous breaking gives rise to a pseudo-Nambu-Goldstone boson known as the axion~\cite{Peccei:1977hh,Weinberg:1977ma,Wilczek:1977pj}. The axion mass and couplings are determined by the PQ symmetry breaking scale $f_a$, while its interaction structure depends on the ultraviolet completion. In the KSVZ model, the couplings of axions to SM fermions are via the heavy beyond SM (BSM) quark loops~\cite{Kim:1979if,Shifman:1979if}, and in the DFSZ model, the axion couples directly to the SM fermions~\cite{Dine:1981rt,Zhitnitsky:1980tq}. Furthermore, the axions can be potentially very heavy, even up to the GeV scale in some scenarios~\cite{Alves:2017avw,Girmohanta:2024nyf,Murayama:2026ioh}. More generally, the axion-like particles (ALPs) share a similar interaction structure without necessarily being tied to the strong CP problem (see e.g. Refs.~\cite{Raffelt:1990yz,Kim:2008hd,Irastorza:2018dyq,DiLuzio:2020wdo,Carenza:2024ehj} for reviews). Both axions and ALPs are well-motivated dark matter (DM) candidates and represent important targets in the search for the BSM physics. For the sake of cleanness, we will use the term ``axion'' for both of them from now on in this paper.

The Sun provides a powerful astrophysical source of axions. The thermal plasma processes such as the Primakoff effect ($\gamma + Ze \to Ze+a$), the electron Compton scattering ($\gamma + e^- \to e^- + a$), and the bremsstrahlung process ($e + X \to e + X + a$ with $X$ referring to an electron or nucleus) produce a continuous spectrum of solar axions at the keV scale
\cite{Redondo:2013wwa,Hoof:2021mld}.
Such solar axions have been extensively searched for by experiments including 
CAST~\cite{CAST:2017uph,CAST:2024eil}, BabyIAXO~\cite{IAXO:2024wss}, RES-NOVA~\cite{Alloni:2026jgy}, NuSTAR~\cite{Ruz:2024gkl}, LUX~\cite{LUX:2017glr}, and PandaX~\cite{PandaX:2017ock}, EDELWEISS~\cite{EDELWEISS:2018tde}, LUX~\cite{LUX:2017glr}, LZ~\cite{LZ:2021xov}, XENON1T~\cite{XENON:2020rca,VanTilburg:2020jvl} and XENONnT~\cite{XENON:2022ltv}. Such solar axions may also generate $X$-rays~\cite{DeRocco:2022jyq,Beaufort:2023zuj} and affect the helioseismology and neutrino observations of the Sun~\cite{Gondolo:2008dd,Vinyoles:2015aba}. 

In addition, nuclear reactions in the solar interior produce monochromatic axion lines~\cite{Raffelt:1982dr,Massarczyk:2021dje,Krcmar:2001si}; see Ref.~\cite{Massarczyk:2021dje} for the full list of nuclear transitions for axion production.
Among these, the proton-deuteron fusion reaction in the proton-proton chain, proceeding via an M1 nuclear transition, emits axions at 5.49\,MeV via
\begin{equation}
\label{eq:reaction}
p + d \to {}^3\mathrm{He} + a \; (5.49\,\mathrm{MeV}), 
\end{equation}
which is one of the most intense monochromatic solar
axion sources \cite{Raffelt:1982dr,Borexino:2012guz}.
Such monochromatic axion production requires hadronic
couplings. The most general couplings of an axion
to nucleons $N$ can be written as,
\begin{equation}
    \mathcal{L}_\mathrm{int} \supset 
    - i \, a\,\bar{N}\gamma_5\left(g_{0aN} + \tau_3 g_{3aN}\right)N.
    \label{eq:nu_lagrangian}
\end{equation}
With $\tau_3$ being the third Pauli matrix acting in the isospin space, $g_{0aN}$ and $g_{3aN}$ are the isoscalar and isovector couplings, respectively. The M1 nuclear transition in \geqn{eq:reaction}
in the Sun is relevant only to the isovector coupling $g_{3aN}$~\cite{Donnelly:1978ty, Raffelt:1982dr,Massarczyk:2021dje}. 
For axions with MeV mass, the most stringent constraints on the couplings $g_{0aN}$ and $g_{3aN}$ are from the supernova observations, which depend to some extent on the axion model details~\cite{Lee:2018lcj,Lella:2023bfb,Lella:2024dmx,Alonso-Gonzalez:2024ems, Alonso-Gonzalez:2024spi, Alonso-Gonzalez:2026syw}. The isovector coupling $g_{3aN}$ is also limited by the axion-induced dissociation of deuterons in  the SNO experiment~\cite{Bhusal:2020bvx} (see  Ref.~\cite{AxionLimits} for a comprehensive list of axion limits).

Although the 5.49\,MeV energy is quite sizable itself,
the scattering with a heavy nucleus such as xenon may
still not generate large enough nuclear recoil to overcome the detection
threshold. For these solar axions to be detected at DM direct detection
experiments, it is necessary to involve the couplings with electrons
($g_{ae}$) or photons ($g_{a \gamma}$),
\begin{equation}
    \mathcal{L}_\mathrm{int} \supset 
    -i \, g_{ae}\, a\, \bar{e}\gamma_5 e 
    - \frac{1}{4}\, g_{a\gamma} \, a F_{\mu\nu}\tilde{F}^{\mu\nu} \,,
    \label{eq:lagrangian}
\end{equation}
where $F_{\mu\nu}$ is the standard electromagnetic field tensor with
$\tilde{F}^{\mu\nu}$ being its dual. The 5.49\,MeV solar axion offers distinctive advantages as a probe in the high-precision terrestrial experiments: the monochromatic signal provides a sharp peak readily distinguishable from the continuous backgrounds. In addition, the detector backgrounds are generally suppressed at MeV energies compared to the keV range at the DM direct detection experiments \cite{PandaX:2014mem}. 
In this paper, we focus on the prospects of the monochromatic 5.49\,MeV solar axions at the xenon experiments such as XENONnT~\cite{XENON:2024wpa}, PandaX-xT~\cite{PANDA-X:2024dlo} and LZ~\cite{LZ:2019sgr}, which are the leading DM direct detection experiments.

For the coupling with electron, we consider the axion mass within the range of $m_a \in [0, 1]~\mathrm{MeV}$ such that the decay $a \to e^+e^-$ is kinematically forbidden. It can induce the $e^+ e^-$ pair production (PP), inverse-Compton (IC) and axio-electric (AE) processes. The signal rates at the detectors depend on the axion mass $m_a$ and the coupling product $g_{3aN} g_{ae}$.  At the xenon target, the PP process is more important than the other two channels, which is largely enhanced by the atomic number squared $Z^2$ \cite{Arias-Aragon:2024gdz}. Furthermore, when the axion mass is nonzero, the corresponding cross section also increases significantly with mass. It is found that the mass enhancement factor can reach ${\cal O} (100)$ at $m_a = 1$\,MeV with respect to the massless case. Benefiting from these two enhancement factors, the sensitivity of $| g_{3aN} g_{ae}|$ can be improved up to $1.59 \times 10^{-14}$ at the 90\% C.L. with an exposure of 200\,ton$\cdot$yr at the xenon experiments. With the ultimate exposure goal of 1000\,ton$\cdot$yr at XLZD~\cite{XLZD:2024nsu}, the prospect can be further pushed to $7.12 \times 10^{-15}$.
For comparison, the most stringent laboratory constraint is currently provided by Borexino: at $m_a=0$, the signals are dominated by the IC process, and the couplings are bound to be  $|g_{3aN} g_{ae}| < 1.90 \times 10^{-13}$ at the 90\% confidence level (C.L.)~\cite{BOREXINO:2025dbp}.

If the axion couples with photons, the signal is the inverse-Primakoff (IP) process with axions converting into monochromatic photons by scattering off the target nuclei. Such a process is also enhanced by the $Z^2$ of xenon. With the fiducial exposures of 200 ton$\cdot$yr and 1000 ton$\cdot$yr, the couplings $|g_{3aN} g_{a\gamma}|$ can be probed up to $6.76\times10^{-12}$ GeV$^{-1}$ and $3.02\times10^{-12}$ GeV$^{-1}$ at the 90\% C.L., respectively, well surpassing the current limit of $2.30\times10^{-11}$ GeV$^{-1}$ by Borexino~\cite{BOREXINO:2025dbp}.

It is remarkable that the prospects of both the couplings $|g_{3aN} g_{ae}|$ and $|g_{3aN} g_{a\gamma}|$ at the next-generation xenon experiments are comparable to or even better than those at the large neutrino detectors such as JUNO~\cite{JUNO:2021vlw} and Hyper-Kamiokande (Hyper-K)~\cite{Hyper-Kamiokande:2018ofw} with much larger exposures. The main driving advantage of the xenon detectors lies in the $Z^2$ enhancement of xenon with respect to carbon in JUNO and oxygen in Hyper-K.

The rest of this paper is organized as follows.
The 5.49\,MeV solar axion flux is sketched in \gsec{sec:flux}.
The sensitivities of $|g_{3aN} g_{ae}|$ at the xenon experiments  
are estimated in \gsec{sec:gae}, and the corresponding prospects of $|g_{3aN} g_{a\gamma}|$ are obtained in \gsec{sec:gay}. In these two sections, we also compare our results with the current limits from Borexino and future prospects at JUNO and Hyper-K. We summarize and conclude in \gsec{sec:conclusion}.

\section{Monochromatic 5.49\,MeV Solar Axion from $pp$ Chain}
\label{sec:flux}

Axions can be produced in the solar interior through nuclear reactions via the axion-nucleon coupling. In particular, the proton-deuteron fusion reaction in the proton-proton chain in \geqn{eq:reaction}
proceeds through an M1 nuclear transition and produces a monochromatic axion flux with energy $E_a = 5.49$\,MeV
\cite{Raffelt:1982dr,Borexino:2012guz,Massarczyk:2021dje,Lucente:2022esm}. For this reaction, the M1 transition is of pure isovector character \cite{Donnelly:1978ty, Raffelt:1982dr,Massarczyk:2021dje}, so that the axion emission rate receives contributions only from the isovector axion-nucleon coupling $g_{3aN}$ in \geqn{eq:lagrangian}~\cite{Lucente:2022esm}. The axion emission rate $\Gamma_a$ relative to the corresponding photon emission rate $\Gamma_\gamma$ from the nuclear fusion $p + d \to {}^3\mathrm{He} + \gamma$ is given by~\cite{Donnelly:1978ty,Raffelt:1982dr,Borexino:2012guz}
\begin{equation}
    \frac{\Gamma_a}{\Gamma_\gamma} \ \simeq \ 0.54\, g_{3aN}^2 \beta_a^3 \,,
    \label{eq:branching}
\end{equation}
where $\beta_a \equiv \sqrt{E_a^2-m_a^2}/E_a$ is the velocity of axion and the factor $\beta_a^3$ accounts for the phase space modification due to a finite axion mass $m_a$. Since the reaction in \geqn{eq:reaction} is the second step of the solar $pp$ chain, its rate is fixed by that of the preceding $p + p \to d + e^+ + \nu_e$ reaction, so that the axion flux is normalized to the standard solar model $pp$ neutrino flux $\Phi_{\nu pp} = 6.0\times10^{10}~\mathrm{cm}^{-2}\mathrm{s}^{-1}$, multiplied by the branching ratio $\Gamma_a/\Gamma_\gamma$~\cite{Bahcall:2004pz}. 
The axion flux at the Earth is then given by~\cite{Raffelt:1982dr,Borexino:2012guz,Lucente:2022esm}
\begin{equation}
    \Phi_a \ = \ \Phi_{a0}\, e^{-d_\odot/\ell_a} 
    \ \simeq \ 3.23 \times 10^{10}\, g_{3aN}^2 \beta_a^3 \,
    e^{-d_\odot/\ell_a} \;  \mathrm{cm}^{-2}\; \mathrm{s}^{-1} \,,
    \label{eq:flux_full}
\end{equation}
where $d_\odot$ is the Sun-Earth distance and $\ell_a$ is the axion decay length in the laboratory frame taking into account the Lorentz boost factor.

In this work, we focus on the axion mass range $m_a \in [0, 1]\,\mathrm{MeV}$, where the decays $a \to e^+e^-$ and $a \to N\bar{N}$ are kinematically forbidden, leaving $a \to \gamma\gamma$ as the only open decay channel. For the axion-photon couplings $g_{a\gamma}$ relevant to this work, the corresponding decay length $\ell_a$ is many orders of magnitude larger than the Sun-Earth distance $d_\odot$~\cite{Qiu:2026irb}. 
As a result, the exponential factor $e^{-d_\odot/\ell_a}$ can be safely neglected, and the axion flux can be simplified to
\begin{equation}
    \Phi_a \ \simeq \ 3.23 \times 10^{10}\, g_{3aN}^2 \beta_a^3 \,
    \mathrm{cm}^{-2} \; \mathrm{s}^{-1} \,.
    \label{eq:flux}
\end{equation}
For the mass range of $m_a \leq 1$\,MeV, the phase space factor satisfies 
$\beta_a^3 \gtrsim 0.93$, and the flux reduction due to the finite axion mass is at the order of a few percent, which is taken into account in our sensitivity calculations below. 

\section{Direct Detection with Axion-Electron Coupling}
\label{sec:gae}

Given the axion coupling with electron, there are multiple detection channels at the xenon detectors, i.e. the axion-induced $e^+ e^-$ PP, IC and AE processes. With a fixed axion energy of 5.49\,MeV, it turns out the detection of axions is dominated by the $e^+ e^-$ PP process (cf.~\gfig{fig:sigma_energy}), which is detailed in~\gsec{sec:pair}. The two subleading channels are sketched in~\gsec{sec:subdominant}. The resultant sensitivities at PandaX are estimated in \gsec{sec:sens_gae}.

\subsection{Axion-induced $e^+ e^-$ pair production}
\label{sec:pair}

\begin{figure}[t]
\centering
\includegraphics[width=0.95\textwidth]{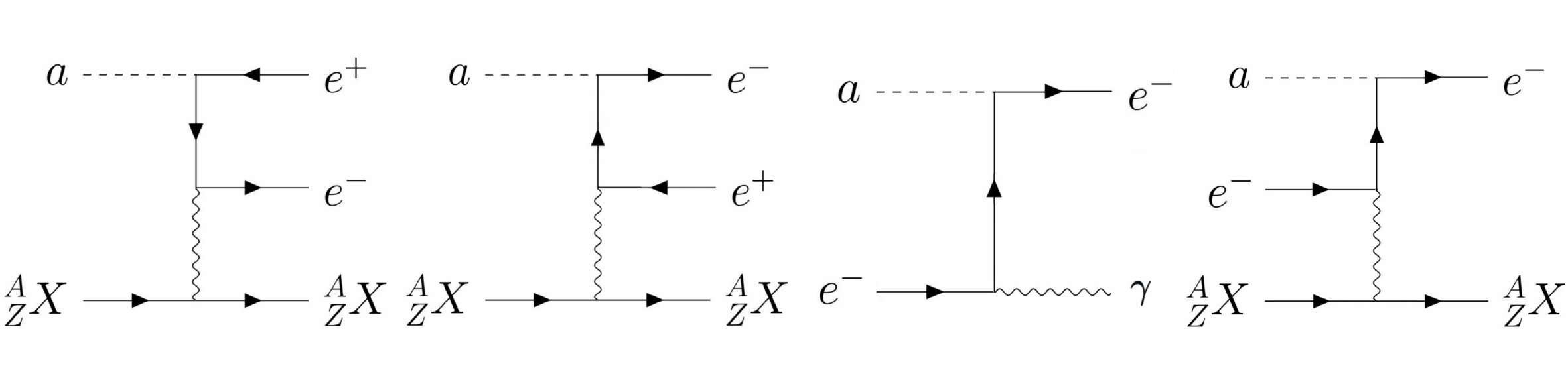}
\caption{Feynman diagrams for the $e^+ e^-$ pair
production (the two left panels), inverse Compton (the third panel)
and axio-electric (the last panel) processes, which
are induced by the axion-electron coupling $g_{ae}$. 
}
\label{fig:feynman}
\end{figure}

An important detection channel for the 5.49\,MeV solar axions in xenon experiments is the axion-induced $e^+ e^-$ pair production (PP) process ($a + {}^A_Z X \to {}^A_Z X + e^+ + e^-$) in the Coulomb field of a nucleus $_Z^A X$~\cite{Bardeen:1978nq,Zhitnitsky:1979cn,Kim:1982xb,Kim:1984ss,Blumlein:1991xh,Arias-Aragon:2024gdz}.
This process is the axion analogue of the Bethe-Heitler PP process induced by a photon, $\gamma + X \to X + e^+ + e^-$~\cite{Bethe:1934za,Motz:1969ti}. 
The two contributing Feynman diagrams, with the axion coupled to the external positron or electron lines, respectively, are shown in the first two panels of \gfig{fig:feynman}. 
The corresponding cross section $\sigma_{\rm PP}(m_a,\,E_a;\, X)$ depends not only on the axion mass $m_a$ and energy $E_a$, but also on the nucleus ${}_Z^A X$ involved.
In the massless limit ($m_a = 0$), the axion-induced PP cross section $\sigma_{\rm PP}$ is related to the cross section $\sigma_{\gamma ee} (E_\gamma;X)$ for the Bethe-Heitler process \cite{Arias-Aragon:2024gdz},
\begin{equation}
    \sigma_{\rm PP}(0,E_a;X) \ \simeq \ 2.6\, \frac{g_{ae}^2}{4\pi\alpha}\, 
    \sigma_{\gamma ee}(E_\gamma = E_a;X) \,,
    \label{eq:pair_relation}
\end{equation}
with $\alpha$ is the fine-structure constant.  
The cross section $\sigma_{\gamma ee}$ can be easily obtained from the NIST XCOM database~\cite{NIST-XCOM}. At $E_a =$ 5.49\,MeV, the axion-induced PP cross section on xenon is
\begin{equation}
    \sigma_{\rm PP}(0,5.49\,{\rm MeV};\mathrm{Xe}) \ = \ 1.01 \times 10^{-22} \, g_{ae}^2~\mathrm{cm}^2 \,.
    \label{eq:sigma_xe}
\end{equation}
In this calculation, we have taken the standard atomic weight $A = 131.293$ for xenon, which corresponds to the natural isotopic abundance average. The differences in the PP cross section among the individual xenon isotopes ($A = 128 \sim 136$) are negligible and would not be discernible in an actual experiment.

\begin{figure}[t]
\centering
\includegraphics[width=0.32\textwidth]{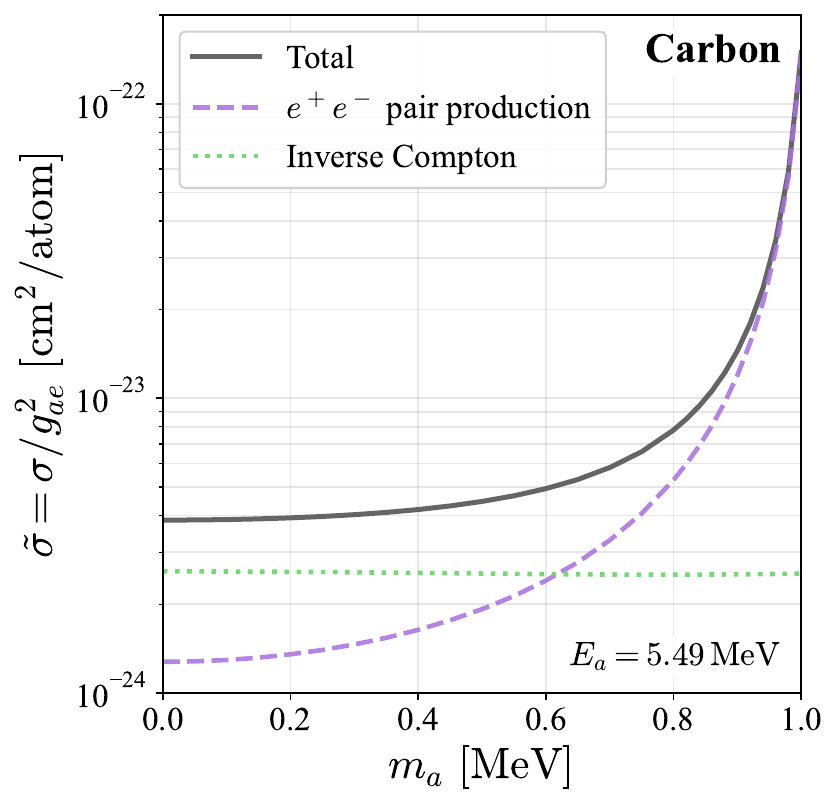}
\includegraphics[width=0.32\textwidth]{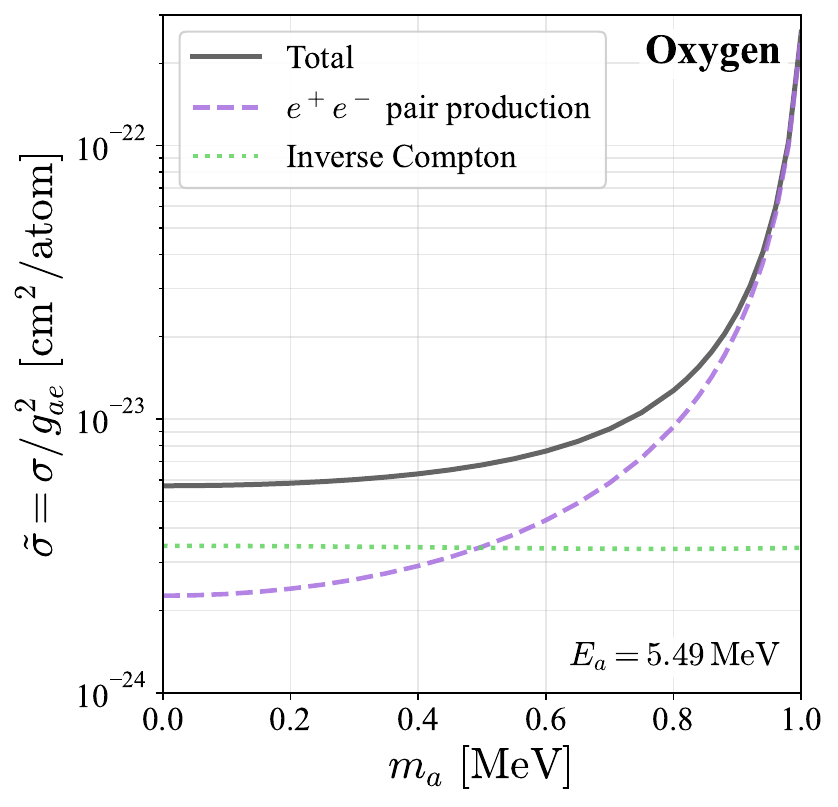} 
\includegraphics[width=0.32\textwidth]{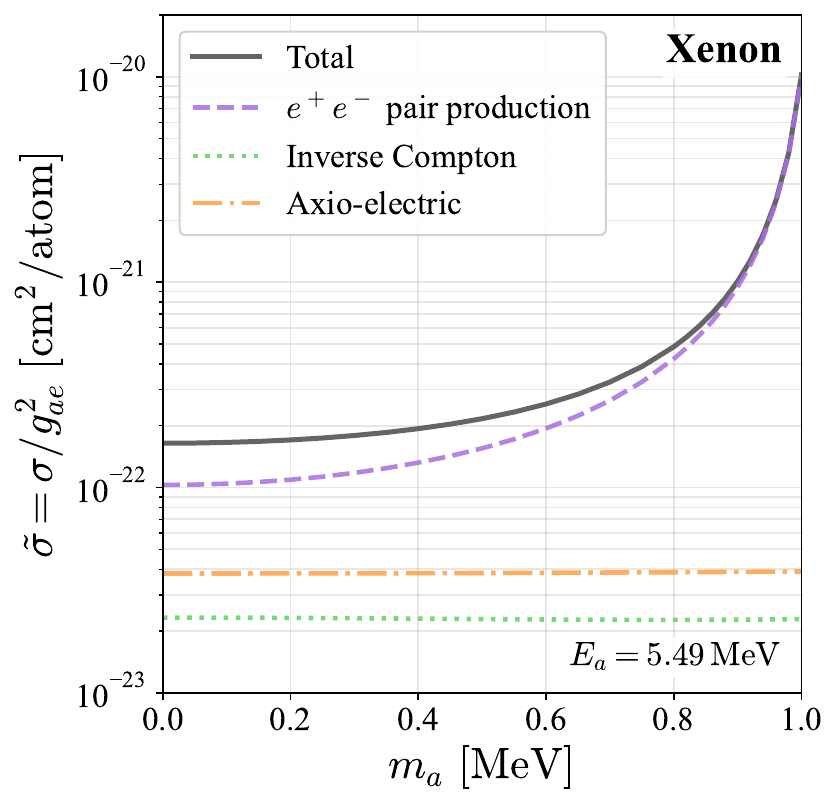}
\caption{Effective cross sections $\tilde{\sigma} \equiv \sigma/g_{ae}^2$ for the $g_{ae}$-induced detection channels of $e^+ e^-$ PP (dashed purple), IC (dotted green), and AE (dash-dotted orange) as functions of the axion mass $m_a$ with fixed axion energy $E_a = 5.49$\,MeV. 
The dark lines are for the total cross sections summing them up. The left, middle, and right panels are for the carbon, oxygen, and xenon targets, respectively. 
Since the AE cross section is much
smaller than those of the other two processes for
carbon and oxygen targets, the corresponding lines are not
shown in the left and middle panels. 
}
\label{fig:sigma_mass}
\end{figure}

The total cross section $\sigma_{\rm PP} (m_a, E_a;{\rm Xe})$ with a non-zero axion mass $m_a$ can be calculated in a straightforward way. To be more specific, the squared matrix elements and phase space integration details can be found in the appendix of Ref.~\cite{Arias-Aragon:2024gdz}.\footnote{We note that the momentum transfer variable $t$ in the nuclear form factor of Ref.~\cite{Arias-Aragon:2024gdz} should be understood as the positive quantity $t = -q^2 > 0$, following the convention of the original reference~\cite{Tsai:1973py}. Using the opposite sign leads to an unphysical divergence of the form factor near $m_a = 1$\,MeV.} Our numerical result reproduces the value in the $m_a = 0$ limit in \geqn{eq:sigma_xe} based on the simple rescaling of the cross section of the photon induced Bethe-Heitler process, which validates our numerical method.
To quantify the effect of a finite axion mass, we define the mass enhancement factor
\begin{equation} 
    R_X (m_a) \ \equiv \ \frac{\sigma_{\rm PP}(m_a, E_a; X)}{\sigma_{\rm PP}(0, E_a; X)} \,,
    \label{eq:R_def}
\end{equation}
with the axion energy fixed at $E_a = $ 5.49\,MeV. In addition to xenon, we consider also carbon and oxygen in this paper, which are the predominant target nuclei of the liquid scintillator detectors Borexino and JUNO as well as the water detector Hyper-K, respectively. The factor $R_X(m_a)$ increases monotonically with the axion mass $m_a$ as shown in \gfig{fig:sigma_mass}. At $m_a = $ 1\,MeV our numerical calculations give
\begin{equation}
    R_{\mathrm{C}}(1\,\mathrm{MeV}) \simeq 115\,, \qquad
    R_{\mathrm{O}}(1\,\mathrm{MeV}) \simeq 113\,, \qquad
    R_{\mathrm{Xe}}(1\,\mathrm{MeV}) \simeq 99\,.
    \label{eq:R_values}
\end{equation}
The mass enhancement reaches a factor of $\sim 100$. With the values being quite close for carbon, oxygen and xenon, the enhancement is a universal effect almost independent of the target nucleus. We have also checked the mass enhancement for hydrogen, which is present in the experiments Borexino, JUNO and Hyper-K, and obtained a comparable enhancement of $R_{\mathrm{H}}(1\,\mathrm{MeV}) \sim 120$. However, suppressed heavily by the small value of $Z = 1$ of hydrogen, the corresponding contributions to the total event rates in these experiments are negligible, and we do not include it in the analysis below. 
Physically, this mass enhancement here is due to the kinematic reason, as the decay $a \to e^+ e^-$ will be kinematically allowed when $m_a > 2m_e = 1.022$\,MeV. This is quite similar to the case of much larger cross section of two-body decays with respect to three-body decays. The effective cross section $\tilde\sigma_{\rm PP} = \sigma_{\rm PP}/g_{ae}^2$ is shown with dashed purple lines in \gfig{fig:sigma_mass}, as a function of the axion mass $m_a$. 
Benefiting from this enhancement, the sensitivities of all such experiments can be substantially improved for a relatively heavy axion, as we will see clearly in \gsec{sec:sens_gae}.

\begin{figure}[t]
\centering
\includegraphics[width=0.32\textwidth]{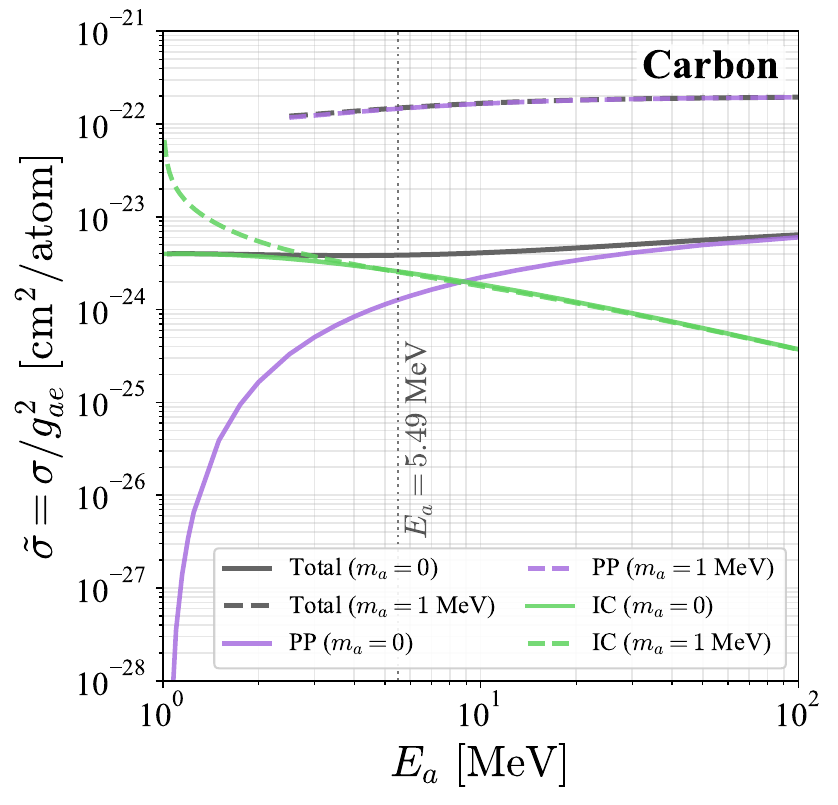}
\includegraphics[width=0.32\textwidth]{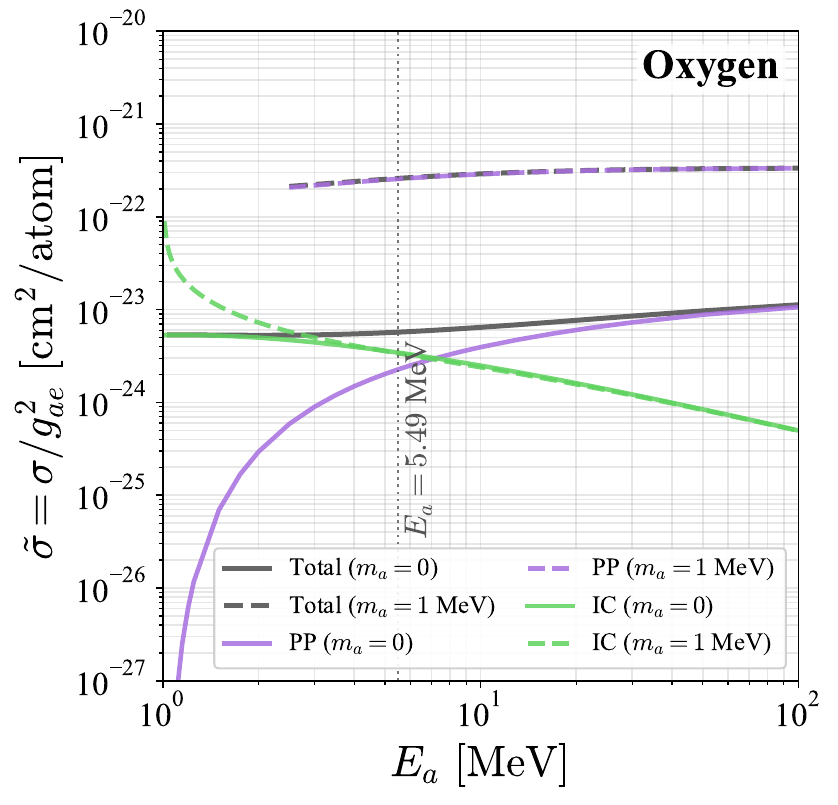} 
\includegraphics[width=0.32\textwidth]{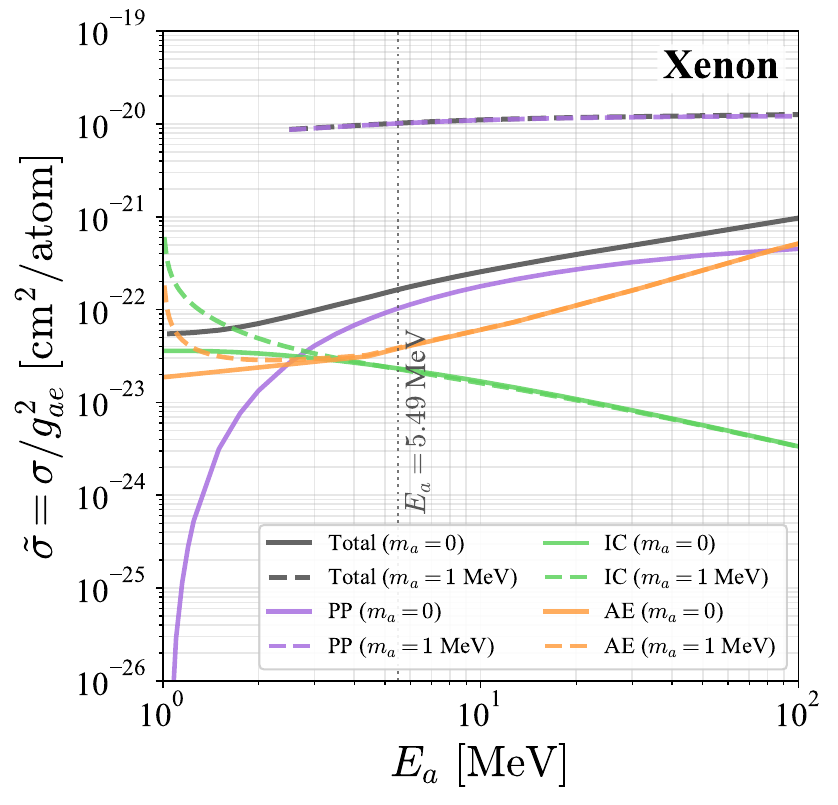}
\caption{
Effective cross sections $\tilde{\sigma} \equiv \sigma/g_{ae}^2$ for the $g_{ae}$-induced detection channels of $e^+ e^-$ PP (purple), IC (green), and AE (orange) as functions of the axion energy $E_a$. The solid and dashed curves correspond to the axion masses $m_a = 0$ and $1$\,MeV, respectively, and the dark lines are for the total cross sections summing all the channels up. The left, middle and right panels are for carbon, oxygen and xenon, respectively. 
Since the AE cross section is much smaller than those of
the other two processes for carbon and oxygen, the
corresponding lines are not shown in the left and middle panels. The fixed solar axion energy of $E_a = 5.49$\,MeV in this paper is indicated by the vertical dotted lines.
}
\label{fig:sigma_energy}
\end{figure}

Since the Bethe-Heitler cross section scales roughly as $\sigma_{\gamma ee} \propto Z^2$~\cite{Bethe:1934za,Motz:1969ti}, \geqn{eq:pair_relation} implies that the cross section for the axion-induced $e^+ e^-$ production process scales also roughly as $\sigma_{\rm PP} \propto Z^2 g_{ae}^2$. This is an important advantage of xenon ($Z=54$) over the relatively light nuclei such as carbon ($Z=6$) and oxygen ($Z=8$). At $E_a = $ 5.49\,MeV,
\begin{subequations}
\begin{align}
    \frac{\sigma_{\rm PP}(0, E_a; \mathrm{Xe})}{\sigma_{\rm PP}(0, E_a; \mathrm{C})} &\simeq 81\,, 
    &\qquad
    \frac{\sigma_{\rm PP}(0, E_a; \mathrm{Xe})}{\sigma_{\rm PP}(0, E_a; \mathrm{O})} &\simeq 46\,, \\[4pt]
    \frac{\sigma_{\rm PP}(1\,\mathrm{MeV}, E_a; \mathrm{Xe})}{\sigma_{\rm PP}(1\,\mathrm{MeV}, E_a; \mathrm{C})} &\simeq 69\,, 
    &\qquad
    \frac{\sigma_{\rm PP}(1\,\mathrm{MeV}, E_a; \mathrm{Xe})}{\sigma_{\rm PP}(1\,\mathrm{MeV}, E_a; \mathrm{O})} &\simeq 40\,.
\end{align}
\label{eq:ratio_xe}
\end{subequations}
For $m_a = 0$, the ratios reproduce the naive $Z^2$ scaling of $(54/6)^2 = 81$ and $(54/8)^2 \simeq 45.6$. For $m_a = 1$\,MeV, the ratios are reduced because the mass enhancement is slightly larger for lighter nuclei $R_{\mathrm{C}}$, $R_{\mathrm{O}} > R_{\mathrm{Xe}}$ (cf. \geqn{eq:R_values} and \gfig{fig:sigma_mass}), partially offsetting the $Z^2$ advantage of xenon. The effective cross section $\tilde\sigma_{\rm PP}$ as a function of the axion energy $E_a$ is shown as purple lines in~\gfig{fig:sigma_energy}. For the case of $m_a = 1$\,MeV, there is a lower threshold for the axion energy, i.e. $E_a \geq m_a + 2m_e \simeq 2.022$\,MeV. 
All the ratios in \geqn{eq:ratio_xe} can be directly obtained from the values of the reduced cross sections at $E_a = 5.49$\,MeV in \gfig{fig:sigma_energy}, indicated by the vertical dashed lines.

\begin{figure}[t]
\centering
\includegraphics[width=0.55\textwidth]{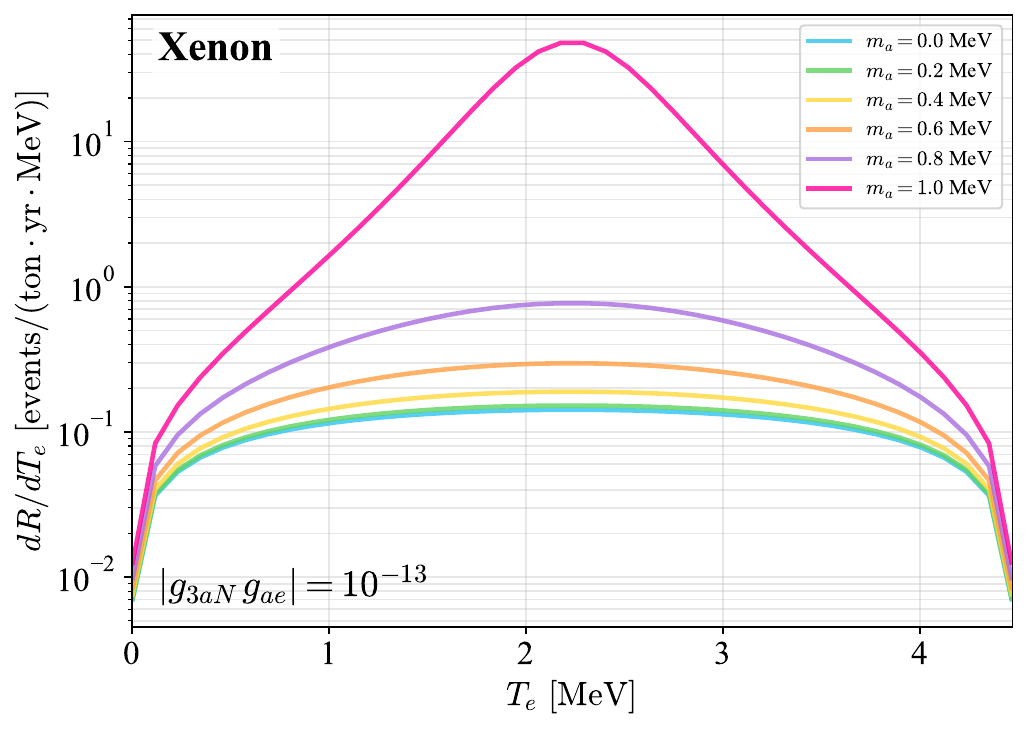}
\caption{Differential event rate ${\rm d}R/{\rm d}T_e$ for
the axion-induced PP process on xenon as a function of the
final-state electron kinetic energy $T_e$ with fixed axion
energy $E_a = 5.49$\,MeV. The axion mass is set to be 0,
0.2, 0.4, 0.6, 0.8 and 1.0\,MeV from bottom to top and we
have set the coupling $|g_{3aN}\,g_{ae}| = 10^{-13}$.
}
\label{fig:spectrum}
\end{figure}

The differential event rate ${\rm d}R/{\rm d}T_e$ of the axion-induced $e^+ e^-$ PP on xenon  
is presented in \gfig{fig:spectrum} as a function of
the final-state electron kinetic energy $T_e \equiv 
E_e - m_e$ (with $E_e$ being the electron energy). We take $E_a = $ 5.49\,MeV and the representative axion masses $m_a = 0,\, 0.2,\, 0.4,\, 0.6,\, 0.8$ and $1.0$\,MeV, as well as $|g_{ae}\,g_{3aN}| = 10^{-13}$ for benchmarks.
Since the $e^+ e^-$ pair in the final state contributes a total mass of $2 m_e = 1.022$\,MeV, the electron kinetic energy can only extend to 4.468\,MeV. 
The spectra are symmetric about $T_e = (E_a - 2m_e)/2 \simeq 2.234$\,MeV around which the enhancement with large axion mass is also more significant. This implies that the $e^+ e^-$ pair in the final state tends to split the kinetic energy equally, in particular for a heavy axion $m_a \sim 1$\,MeV. For the scattering of axions with other nuclei such as carbon and oxygen, the differential spectra are expected to exhibit very similar behavior.

\subsection{Inverse-Compton and axio-electric processes}
\label{sec:subdominant}

Besides the leading process of axion-induced $e^+ e^-$ PP,
there are two subleading detection channels at xenon detectors,
i.e. the IC and AE processes, from the same axion-electron
coupling $g_{ae}$. The corresponding Feynman diagrams are
shown in the third and last diagrams of
\gfig{fig:feynman}, respectively.

In the IC scattering
  $a + e^- \to e^- + \gamma$,
the axion converts into a photon by scattering off an atomic electron. The total cross section is~\cite{Avignone:1988bv}
\begin{align}
  \sigma_{\rm IC}
& =
  \frac{\alpha g_{ae}^2}{8m_e^2|\bm k_a|}
\left[
  \frac{2m_e^2(m_e + E_a)y}{(m_e^2 + y)^2}
+ \frac{4m_e(m_a^4 + 2m_a^2m_e^2 - 4m_e^2E_a^2)}{y(m_e^2 + y)} \right.
\nonumber \\
& \left.
\hspace{21mm}
+ \frac{4m_e^2|\bm{k}_a|^2 + m_a^4}{|\bm{k}_a|y}
  \log \left( \frac{m_e + E_a + |\bm{k}_a|}{m_e + E_a - |\bm{k}_a|} \right)
\right] \,,
\label{eq:sigma_ic}
\end{align}
where $|\bm k_a| = \sqrt{E_a^2 - m_a^2}$ is the
axion momentum and $y \equiv 2m_eE_a + m_a^2$.
In the massless limit $m_a = 0$ at $E_a = 5.49\,\mathrm{MeV}$, this reduces to $\sigma_{\rm IC} \simeq 4.3 \times 10^{-25}\,g_{ae}^2~\mathrm{cm}^2$ per electron.
Since the target for this process is an atomic
electron, the cross section scales linearly with
the xenon atomic number $Z = 54$.

The AE process ($a + e^- + {}^A_ZX \to e^- + {}^A_ZX$) 
is the axion analogue of the photoelectric (PE) effect, with the initial electron being in an atomic bound state. The corresponding cross section $\sigma_{\rm AE}$ is related to the PE cross section $\sigma_{\rm PE}$ 
by~\cite{CUORE:2012ymr}
\begin{equation}
    \sigma_{\rm AE} = \sigma_{\rm PE}\,\frac{g_{ae}^2}{\beta_a}
    \frac{3E_a^2}{16\pi\alpha m_e^2}
    \left(1 - \frac{\beta_a^{2/3}}{3}\right) \,,
    \label{eq:sigma_ae}
\end{equation}
where $\sigma_{\rm PE}$ can be taken directly from the NIST XCOM database~\cite{NIST-XCOM}. 
Notably, the PE cross section scales as $\sigma_{\rm PE} \propto Z^5$.
So the AE effect is largely enhanced for high-$Z$ targets,
in particular when the axion energy is large. Nevertheless,
it remains a subleading channel for our interest in this paper
\cite{Arisaka:2012pb}. 

The (effective) cross sections $\sigma_{\rm IC}$ ($\tilde\sigma_{\rm IC} \equiv \sigma_{\rm IC}/g_{ae}^2$) and $\sigma_{\rm PE}$ ($\tilde\sigma_{\rm PE} \equiv \sigma_{\rm PE}/g_{ae}^2$) do not change too much within the axion mass range of $m_a \in [0,\,1]$\,MeV for the fixed energy of $E_a = 5.49$\,MeV in this paper, as indicated by the dotted green (IC) and orange (AE) lines in \gfig{fig:sigma_mass}. 
Their dependence on the axion energy $E_a$ is presented again as the green and orange lines in \gfig{fig:sigma_energy}, with the solid and dashed lines for $m_a = 0$ and 1\,MeV, respectively, the same as that for the $e^+ e^-$ PP channel. 
For the low-$Z$ carbon and oxygen targets, the AE cross section is much smaller than those of the PP and IC channels, and can be safely neglected. The dash-dotted orange AE line is therefore shown only for xenon, where the $Z^5$ scaling of $\sigma_{\rm PE}$ makes this channel non-negligible. 
In both \gfig{fig:sigma_mass} and \gfig{fig:sigma_energy}, the dark gray lines denote the total effective cross section, summing over all the three detection channels for xenon and over the $e^+ e^-$ PP and IC channels for carbon and oxygen. Here we can compare the production channels for the representative elements carbon, oxygen and xenon.

\begin{itemize}
    \item For the lighter carbon and oxygen targets, the IC process is the dominant process when the axion is light and the axion energy is small. When the axion is heavy, say $m_a \gtrsim 0.6$\,MeV for fixed $E_a = 5.49$\,MeV, or the axion energy becomes larger, say $E_a \gtrsim {\cal O} (10)$\,MeV for $m_a = 0$, the PP channel takes over to be the leading one, as implied by the two left panels of \gfig{fig:sigma_mass} and \gfig{fig:sigma_energy}. 
    Throughout the parameter space of interest in this paper, the cross section of the AE channel is much smaller than that of the IC and PP channels for the low-$Z$ carbon and oxygen targets. Therefore, the AE channel can be neglected for these two targets, both in \gfig{fig:sigma_mass} and \gfig{fig:sigma_energy} as well as the sensitivity calculations below.
    
    \item At xenon targets, the IC and AE channels are important for $m_a \sim 0$ and $E_a \lesssim 2$\,MeV, as seen in the right panel of \gfig{fig:sigma_energy}. However, for $E_a \gtrsim$ a few MeV, in particular at $E_a = 5.49$\,MeV, the cross section for the $e^+ e^-$ PP channel is significantly larger than that for the other two channels. For $m_a = 0$, the IC and AE processes both give non-negligible contributions, smaller than the PP channel by a factor of 4.5 and 2.6, respectively.  At $m_a = 1$\,MeV, the mass enhancement of the PP channel makes the total cross section almost entirely determined by this channel. 
\end{itemize}

\subsection{Signal rates and sensitivities}
\label{sec:sens_gae}

For a given detection channel, the expected number $S$ of signal events is
\begin{equation}
  S
\equiv
  \epsilon \cdot N_T \cdot T \cdot \Phi_a \cdot \sigma \,,
\label{eq:signal}
\end{equation}
where $\epsilon$ is the detection efficiency, $N_T$ is the number of target particles, $T$ is the exposure time, $\Phi_a$ is the axion flux given in \geqn{eq:flux}, and $\sigma$ is the relevant detection cross section. Depending on the channels involved, the target is either an electron or a nucleus. For the IC scattering we use the per-electron cross section together with $N_T \equiv Z\,N_{\rm atom}$ where $N_{\rm atom}$ is the number of target atoms, while for the PP and the AE processes the cross sections are defined per atom (nucleus) such that $N_T \equiv N_{\rm atom}$.

Since the axion flux scales as $\Phi_a \propto |g_{3aN}|^2$ and each detection cross section scales as $|g_{ae}|^2$, the number of signal events is proportional to the square of the coupling product $|g_{3aN} g_{ae}|^2$. Combining the contributions from different channels, the total signal reads
\begin{equation}
  S_{\rm total}
\equiv
  \sum_{i} S_i,
\qquad \text{with} \qquad
  S_i
\equiv
  K_{i} \, g_{ae}^2 g_{3aN}^2, 
\label{eq:Stotal}
\end{equation}
where $S_i$ refers to the number of signal events
in the $i$-th detection channel. In addition,
$K_i$ is the proportionality factor which collects
the corresponding efficiency, target number,
exposure, flux, and the effective cross section
(excluding the couplings) in \geqn{eq:signal}.
For a given upper limit $S_{\rm lim}$ on the number of signal events at a certain C.L., the combined sensitivity on $|g_{3aN} g_{ae}|$ follows from setting $S_{\rm total} = S_{\rm lim}$, which gives $|g_{3aN} g_{ae}| = \sqrt{S_{\rm lim}/K_{\rm total}}$ with $K_{\rm total} \equiv \sum_i K_i$. The sensitivity of an individual channel follows similarly by setting $S_i = S_{\rm lim}$. 
We are now ready to obtain the recast Borexino limit and the prospects at the xenon experiments, JUNO and Hyper-K.

\paragraph{Xenon experiments:} 
In this work, we consider three benchmark configurations for the xenon experiments. 
\begin{itemize}
  \item The (neutrinoless) double-beta decay signals of ${}^{136}{\rm Xe}$ are at the few MeV scale, and the corresponding fiducial volume at PandaX-4T has 3.7\,tonne of natural xenon \cite{PandaX:2022kwg}. Based on the PandaX-4T experiment, we take the first benchmark configuration to be $7.4\,\mathrm{ton}\cdot\mathrm{yr}$, which is equivalent to two years of physics data taking.
    
  \item The future PandaX-xT experiment is projected to achieve a total exposure of $200\,\mathrm{ton}\cdot\mathrm{yr}$ which is adopted as our second benchmark configuration \cite{PANDA-X:2024dlo}.

  \item The XLZD is a proposed next-generation liquid xenon experiment jointly pursued by the XENON, LUX-ZEPLIN, and DARWIN collaborations, with a baseline target of $60 \sim 80$\,tonnes of active xenon and an exposure up to 1000\,ton$\cdot$yr \cite{XLZD:2024nsu}. Following XLZD, we take the third benchmark value for the exposure to be 1000\,ton$\cdot$yr.
\end{itemize}

\begin{figure}[t]
\centering
\includegraphics[width=0.7\textwidth]{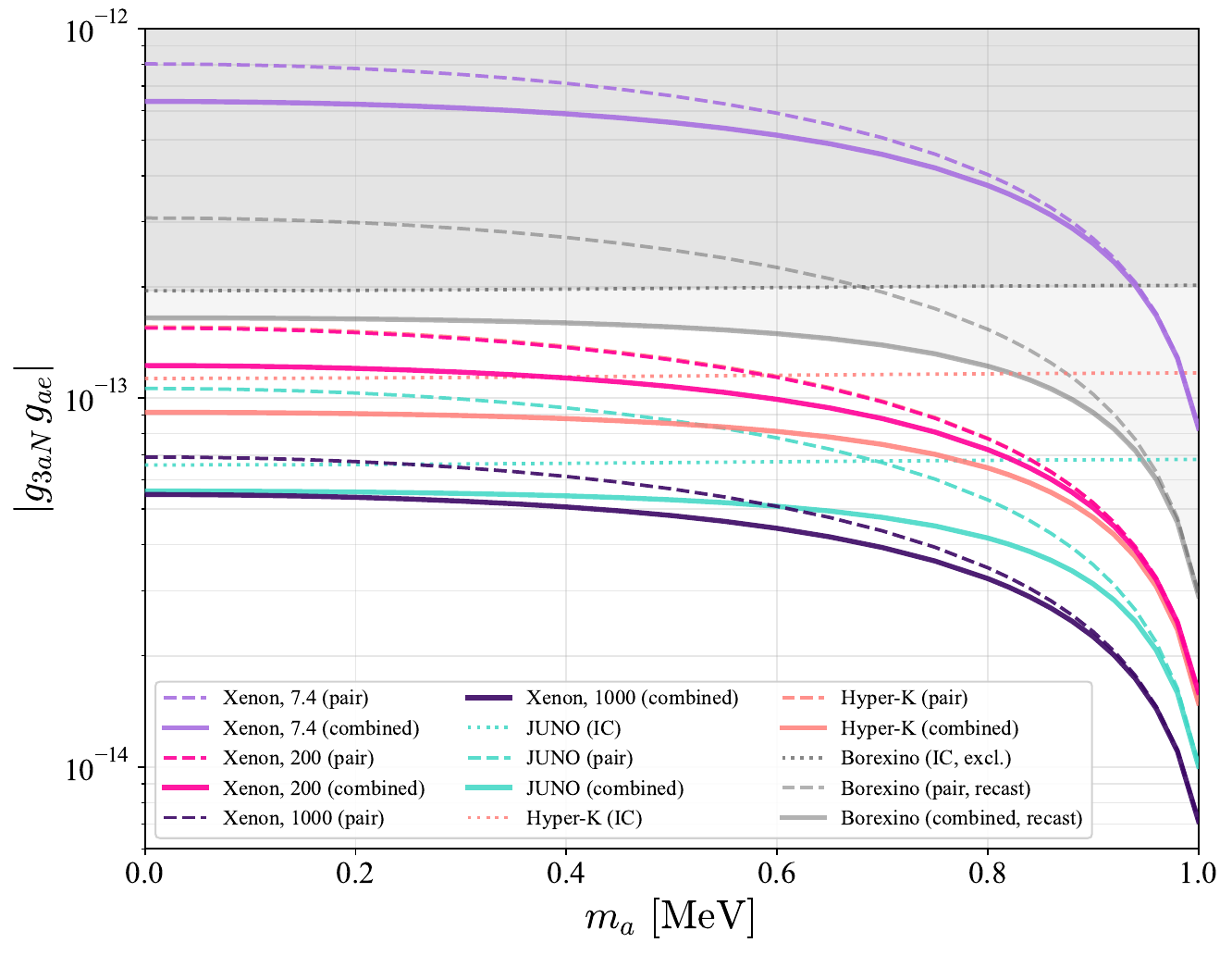}
\caption{The 90\%\,C.L. sensitivities of $|g_{3aN} g_{ae}|$ at xenon experiments with exposures of $7.4\,\mathrm{ton}\cdot\mathrm{yr}$ (light purple), $200\,\mathrm{ton}\cdot\mathrm{yr}$ (magenta) and $1000\,\mathrm{ton}\cdot\mathrm{yr}$ (dark purple) as functions of the axion mass $m_a$. Also shown are the recast Borexino limits (shaded gray)~\cite{BOREXINO:2025dbp}, and the projections for JUNO ($1.62\times10^5\,\mathrm{ton}\cdot\mathrm{yr}$, light green) and Hyper-K ($1.87\times10^6\,\mathrm{ton}\cdot\mathrm{yr}$, light orange)~\cite{Arias-Aragon:2024gdz}. The dashed and dotted lines are for the PP and IC channels, respectively, and the solid lines are for the combined limits/prospects. The main results have been collected in Table\,\ref{tab:comparison_gae}.
}
\label{fig:sens_gae}
\end{figure}

For all the three benchmark configurations above, we assume a uniform detection efficiency of $\epsilon = 100\%$ in all the detection channels~\cite{PANDA-X:2024dlo}. Note that liquid xenon detectors operate with extremely low backgrounds in the MeV energy range.
Although the precise background rate around $5.49$\,MeV requires a dedicated experimental analysis, the very low background together with the monochromatic signal motivates a zero-background approximation. Under this assumption, the $90\%$\,C.L. upper limit on the number of signal events corresponds to the Poisson probability $e^{-\mu} = 0.1$, giving $S_\mathrm{lim} = 2.3$. 
The resulting $90\%$\,C.L. sensitivities on $|g_{3aN} g_{ae}|$ with the exposures of 7.4, 200, and 1000\,ton$\cdot$yr are shown as the light purple, magenta and dark purple lines in \gfig{fig:sens_gae}, respectively, with the PP only and the combined three-channel sensitivities depicted as the dashed and solid lines. 
At $m_a = 0$, the combined sensitivity reaches $|g_{3aN} g_{ae}| \lesssim 6.36 \times 10^{-13}$, $1.22 \times 10^{-13}$ and $5.47 \times 10^{-14}$, respectively, for the three benchmark values of exposure above. As a result of the mass enhancement (cf.~\geqn{eq:R_values}) at $m_a = 1$\,MeV, these sensitivities are improved up to $ 8.28 \times 10^{-14}$, $1.59 \times 10^{-14}$ and $7.12 \times 10^{-15}$, respectively.

\paragraph{Borexino:} The monochromatic 5.49\,MeV solar axions have been searched for at some other experiments, e.g. the early experiment with Bi$_4$Ge$_3$O$_{12}$ bolometric detectors, but the limit is rather weak $|g_{ae}\, g_{3aN}| < 1.9\times10^{-10}$ at $m_a=0$~\cite{Derbin:2013zba,Derbin:2014xzr}. 
The strongest current experimental limit on the $g_{ae}$ coupling to $5.49$\,MeV solar axions comes from Borexino, a carbon-based liquid scintillator ($\mathrm{C_9H_{12}}$) detector~\cite{Borexino:2012guz,BOREXINO:2025dbp}. In this target, the signal is dominated by the IC process when the axion is light (cf. \gfig{fig:sigma_mass}). 
Using the complete Borexino dataset ($3995$~days of live-time), the total statistics is $N_e T = 579.3\,\mathrm{kton\cdot days}$, and the detection efficiency in the IC channel is $\epsilon_{\rm IC} = 0.251$\,\cite{BOREXINO:2025dbp}. The $90\%$\,C.L. upper limit on the number of registered signal events is $S_\mathrm{lim}^{\rm IC} = 8.7$, which already includes the detection efficiency. The sensitivity then follows from $\Phi_a\,\sigma_{\rm IC}\,N_e T \le S_\mathrm{lim}^{\rm IC}$, giving $|g_{3aN} g_{ae}| \le 1.90 \times 10^{-13}$ which is in agreement with the published Borexino value~\cite{BOREXINO:2025dbp}. Extending the axion mass up to $1$\,MeV, this limit is almost a constant, varying from $1.90 \times 10^{-13}$ at $m_a = 0$ to $\sim 2.00 \times 10^{-13}$ at $m_a = 1$\,MeV, since both the IC cross section in \geqn{eq:sigma_ic} and the axion flux in \geqn{eq:flux} depend very weakly on $m_a$ for the mass range of $[0,\,1]$\,MeV. It is shown in \gfig{fig:sens_gae} as the shaded region above the dotted gray line.

For a self-consistent comparison with the xenon prospects, we note that the same mass enhancement also applies to the $e^+e^-$ PP channel on carbon (cf. \geqn{eq:R_values}). We therefore recast the Borexino limit to include the PP channel, computed on carbon with the same statistics $N_e T$ and using the limit of $S_\mathrm{lim}^{\rm IC}$. The constraints for the PP channel only and the combined one are presented as the dashed and solid gray lines in \gfig{fig:sens_gae}, respectively.
As seen in this figure, the recast combined Borexino limit is improved from $1.65 \times 10^{-13}$ at $m_a = 0$ to $\sim 2.90 \times 10^{-14}$ at $m_a = 1$\,MeV, benefiting from the mass enhancement in the PP channel. 

\paragraph{JUNO and Hyper-K:} As future projections, we also compute the sensitivities of two large-volume neutrino experiments within the same framework. JUNO uses a liquid scintillator ($\mathrm{C_{19}H_{32}}$) target and Hyper-K uses water ($\mathrm{H_2O}$). Adopting the detector parameters of Ref.~\cite{Arias-Aragon:2024gdz}, we take exposures of $1.62\times10^{5}\,\mathrm{ton}\cdot\mathrm{yr}$ and the limit of $S_\mathrm{lim} = 104$ for JUNO, and the exposure of $1.87\times10^{6}\,\mathrm{ton}\cdot\mathrm{yr}$ and the limit of $S_\mathrm{lim} = 3500$ for Hyper-K, respectively, with a detection efficiency of $\sim 100\%$ for both. 
The projected sensitivities for JUNO and Hyper-K are shown as the light green and light orange lines in \gfig{fig:sens_gae}. Besides, the sensitivity reaches by the IC, PP and combined channels are presented as the dotted, dashed and solid lines, respectively.  
At $m_a = 0$, the combined sensitivities $5.59 \times 10^{-14}$ at JUNO and $9.14 \times 10^{-14}$ at Hyper-K are consistent with \cite{Arias-Aragon:2024gdz}. As a result of the mass enhancement in the PP channel, these sensitivities at $m_a = 1$\,MeV improve to $1.00 \times 10^{-14}$ at JUNO and $1.49 \times 10^{-14}$ at Hyper-K, respectively.

Based on the sensitivity lines in \gfig{fig:sens_gae}, let us compare the current (recast) limits from Borexino and the prospects at the xenon experiments, JUNO and Hyper-K.
\begin{itemize}
    \item With the exposure of 7.4\,ton$\cdot$yr, the xenon experiment sensitivities have been precluded by the recast Borexino limits. 

    \item When the exposure is enhanced to 200\,ton$\cdot$yr, it is expected that the xenon experiments can surpass the recast Borexino limit across the entire mass range of $m_a \in [0, 1]$\,MeV. This is largely due to the high-$Z$ enhancement of the cross sections in xenon with respect to carbon in Borexino, despite the much smaller target mass in the xenon experiments.     
    In particular, the xenon experiments can improve the recast (combined) Borexino experimental limit by a factor of $1.4$ and $1.8$ at $m_a = 0$ and 1\,MeV, respectively. 

    \item The far-future xenon experiments with the fiducial exposure of $1000\,\mathrm{ton}\cdot\mathrm{yr}$ would improve the sensitivities further, reaching $|g_{3aN} g_{ae}| \simeq 5.47 \times 10^{-14}$ and $7.12 \times 10^{-15}$ at $m_a = 0$ and 1\,MeV, respectively. 
    These ultimate xenon prospects are better than the corresponding Hyper-K sensitivities by a corresponding factor of $1.7$ and $2.1$ at $m_a = 0$ and $1$\,MeV. They are also comparable to the JUNO projection at $m_a = 0$ and even better by a factor of $\sim 1.4$ at $m_a = 1$\,MeV, although the exposure of the xenon experiments is smaller than that of JUNO by more than two orders of magnitude.
    
\end{itemize}
All these limits and sensitivities of these experiments at $m_a = 0$ and $1$\,MeV are summarized in \gtab{tab:comparison_gae}.

\begin{table}[!t]
\centering
\caption{The 90\%\,C.L. sensitivities of xenon experiments to the 5.49\,MeV solar axions through the couplings $|g_{3aN} g_{ae}|$, with axion mass $m_a =0$ or $1$\,MeV. Also shown are the current limits from Borexino~\cite{BOREXINO:2025dbp} and the prospects at JUNO and Hyper-K~\cite{Arias-Aragon:2024gdz}. 
Here PP, IC and AE denote the axion-induced $e^+ e^-$ pair production, the inverse-Compton scattering and the axio-electric effect, respectively. See text and Fig.\,\ref{fig:sens_gae} for more details.} \vspace{5pt}
\label{tab:comparison_gae}
\resizebox{\textwidth}{!}{
\begin{tabular}{c|cccccc}
\hline
\multirow{2}{*}{} & \multirow{2}{*}{Target} & Exposure & Efficiency & \multirow{2}{*}{Channels} & $|g_{3aN}g_{ae}|$  & $|g_{3aN}g_{ae}|$ \\
           &        & (ton$\times$yr) & (\%) &         & at $m_a=0$   & at $m_a\simeq1$\,MeV \\
\hline
Borexino & $\mathrm{C_9H_{12}}$ & 1587 & $22.5$--$25.1$ & IC(+PP) & $1.90\times10^{-13}$~\cite{BOREXINO:2025dbp} & $2.91\times10^{-14}$ \\
JUNO & $\mathrm{C_{19}H_{32}}$ & $1.62\times10^5$ & $\sim100$ & PP+IC & $5.499\times10^{-14}$~\cite{Arias-Aragon:2024gdz} & $1.00\times10^{-14}$\\
Hyper-K & $\mathrm{H_2O}$ & $1.87\times10^6$ & $\sim100$ & PP+IC & $9.14\times10^{-14}$~\cite{Arias-Aragon:2024gdz} & $1.49\times10^{-14}$\\
\hline
\multirow{3}{*}{\makecell{ xenon \\ experiments } } & \multirow{3}{*}{Xe} & 7.4 & \multirow{3}{*}{$\sim100$} & \multirow{3}{*}{PP+IC+AE} & $6.36\times10^{-13}$ & $8.28\times10^{-14}$\\
& & 200 & & & $1.22\times10^{-13}$ & $1.59\times10^{-14}$\\
& & 1000 & & & $5.47\times10^{-14}$ & $7.12\times10^{-15}$\\
\hline
\end{tabular}}
\end{table}

The prospects at the future xenon experiments and the neutrino detectors in \gfig{fig:sens_gae} can be applied directly to the specific QCD axion models, for instance the DFSZ axion model with the axion mass and couplings correlated with each other \cite{Dine:1981rt,Zhitnitsky:1980tq}. For the DFSZ-I and DFSZ-II scenarios~\cite{Dine:1981rt,Zhitnitsky:1980tq,DiLuzio:2020wdo}, the prospects at the xenon experiments with the exposure of 200\,(1000)\,ton$\cdot$yr correspond to the axion mass ranges of $0.27 \sim 1.9$\,keV ($0.18 \sim 1.3$\,keV) and $0.48 \sim 26$\,keV ($0.33 \sim 18$\,keV), respectively (cf. Fig.\,4 of Ref.\,\cite{Arias-Aragon:2024gdz}).

\subsection{Discussions}

It is meaningful to compare our estimates of the xenon experiments with some other constraints on the axion couplings with nucleon and electron. Based on this, we would also clarify some of the differences of the couplings involved. 
\begin{itemize}
    \item In addition to the 5.49\,MeV axion from nucleon fusion, other solar axion sources are also of great interest, e.g. axions from the 14.4\,keV M1 de-excitation transition of $^{57}\mathrm{Fe}^*\to{}^{57}\mathrm{Fe}+a$. The searches of the monochromatic 14.4\,keV signals have been performed in some of the DM direct detection experiments via the AE effect, e.g. CUORE~\cite{CUORE:2012ymr},  MAJORANA~\cite{Majorana:2016hop}, EDELWEISS~\cite{Armengaud:2013rta,EDELWEISS:2018tde}, CDEX~\cite{CDEX:2016rpr,CDEX:2019exx},  XENON1T~\cite{XENON:2020rca} and PandaX-4T~\cite{PandaX:2024cic}. The most stringent limit of $2.07\times10^{-18}$ on the coupling combination $|g_{ae}g_{aN}^{\rm eff}|$ is obtained from a recent reanalysis of the public XENONnT  dataset~\cite{Geng:2025yoj}. Here $g_{aN}^{\rm eff}$ is a model dependent axion-nucleon coupling, given by a model-specific combination of $g_{0aN}$ and $g_{3aN}$. Note that $g_{aN}^{\rm eff}$ does not directly correspond to the pure isovector coupling $g_{3aN}$ in this work.

    \item Solar axions can also be searched for via the resonant absorption by a target nucleus, $a + N \to N^* \to N + \gamma$, where the axion is absorbed and a characteristic photon is subsequently emitted. Using the M1 transition of $^{169}$Tm at 8.41\,keV, the limit of $|g_{ae}(g_{0aN}+g_{3aN})| \le 2.81\times10^{-16}$ at 90\%~C.L. is obtained in a cryogenic bolometer experiment based on a $\mathrm{Tm_3Al_5O_{12}}$ crystal~\cite{Abdelhameed:2020hys}. Here the coupling constrained is the sum $g_{0aN}+g_{3aN}$, and the limit from this experiment can not be directly compared with our results.

    \item Axions can be copiously produced in supernova cores via the nucleon bremsstrahlung $N + N \to N + N + a$~\cite{Carenza:2019pxu} or the pion-induced scattering $\pi^- + p \to n + a$~\cite{Turner:1991ax,Raffelt:1993ix,Keil:1996ju,Carenza:2020cis}. For a heavy axion with mass $m_{a} > 2m_e$, the decay $a \to e^+ e^-$ contributes to the astrophysical 511\,keV lines. As a result, some regions of the parameter space of $g_{ap}=g_{0aN}-g_{3aN}$ and $g_{ae}$ are excluded by the axion-induced energy deposition inside the supernova envelope and the cosmic X-ray background from (extra)galactic supernovae for $m_a > 2m_e$~\cite{Calore:2021klc,Lella:2022uwi}. Here the coupling $g_{an}$ is set to be zero, inspired by the KSVZ axion model~\cite{GrillidiCortona:2015jxo}. These supernova limits can not be directly compared to our results in this paper, and are listed here for completeness.     

    These energetic supernova axions can also be searched for in the neutrino detectors such as JUNO and Hyper-K. It is found that with a galactic supernova explosion, e.g. Betelgeuse at $\sim 200$\,pc, the sensitivities of $|g_{ap} g_{ae}|$ can be improved up to $10^{-20} \sim 10^{-21}$ at JUNO and Hyper-K \cite{Arias-Aragon:2024gdz}. In principle, these supernova limits and the future prospects at neutrino experiments can be applied to more general axion-nucleon couplings, even for the case with $m_a < 2m_e$, which is, however, far beyond the main scope of this paper.
       
\end{itemize}

On the coupling $g_{ae}$ of MeV-scale axions, there are also cosmological constraints from the observations of cosmic microwave background (CMB) and Big Bang nucleosynthesis (BBN)~\cite{Depta:2020zbh,Langhoff:2022bij,Montefalcone:2025nmm}, as well as the  bounds from the X-rays and $\gamma$-rays from the decays $a \to e^+ e^- \gamma$~\cite{Calore:2022pks} and $a\to \gamma\gamma$~\cite{Ferreira:2022egk}. The astrophysical limits on $g_{ae}$ are mainly from the SN1987A data~\cite{Calibbi:2020jvd,Carenza:2021pcm,Calore:2021klc,Ferreira:2022xlw,Fiorillo:2025sln,Ferreira:2025qui} and the most stringent laboratory bounds are set by the experiments PandaX~\cite{PandaX:2024sds}, GERDA~\cite{GERDA:2020emj,GERDA:2024gip}, COSINE-100~\cite{COSINE-100:2023dir} and the MiniBooNE beam-dump~\cite{Capozzi:2023ffu}.

\section{Direct Detection with Axion-Photon Coupling}
\label{sec:gay}

\subsection{Inverse Primakoff process}
\label{sec:primakoff}

\begin{figure}[t]
\centering
\includegraphics[width=0.35\textwidth]{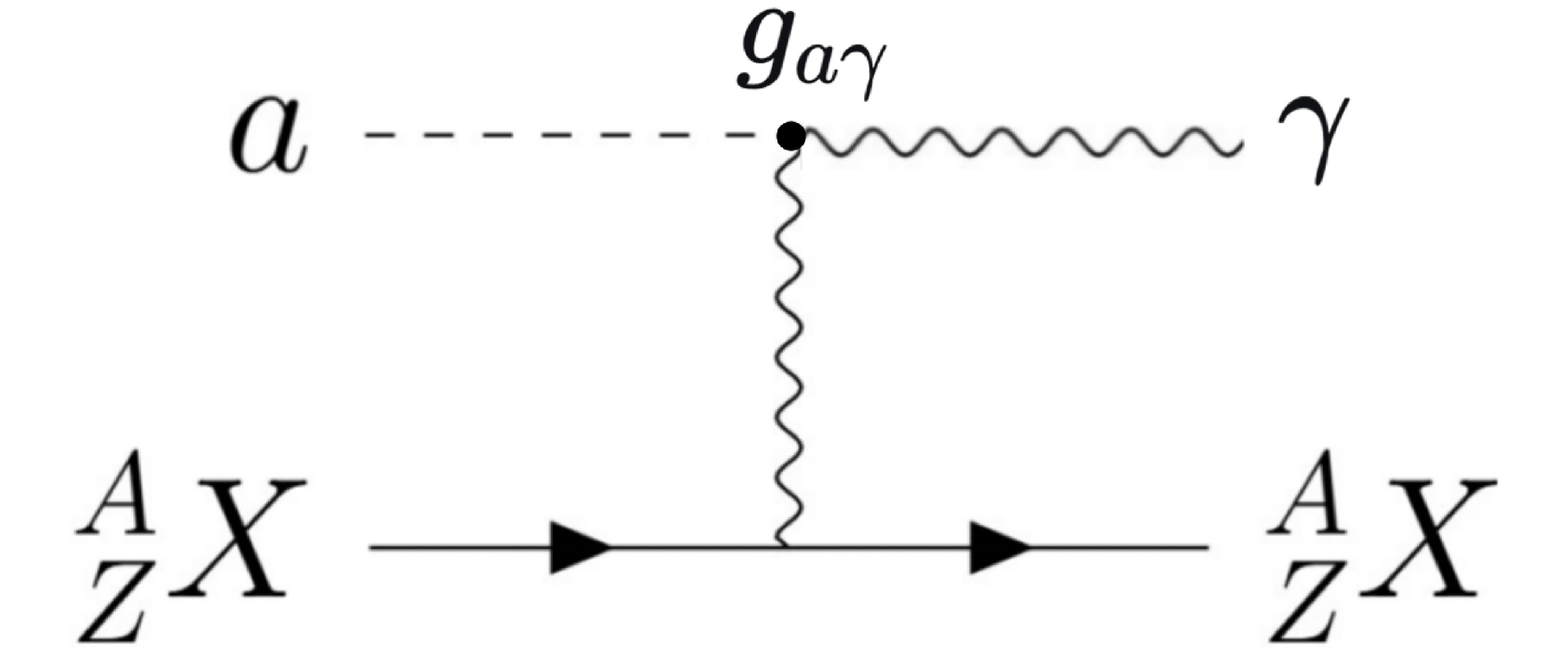}
\caption{Feynman diagram for the IP process
$a + {}^A_Z X \to \gamma + {}^A_Z X$ induced by the axion-photon
coupling $g_{a\gamma}$.}
\label{fig:primakoff_feynman}
\end{figure}

\begin{figure}[t]
\centering
\includegraphics[height=0.35\textwidth]{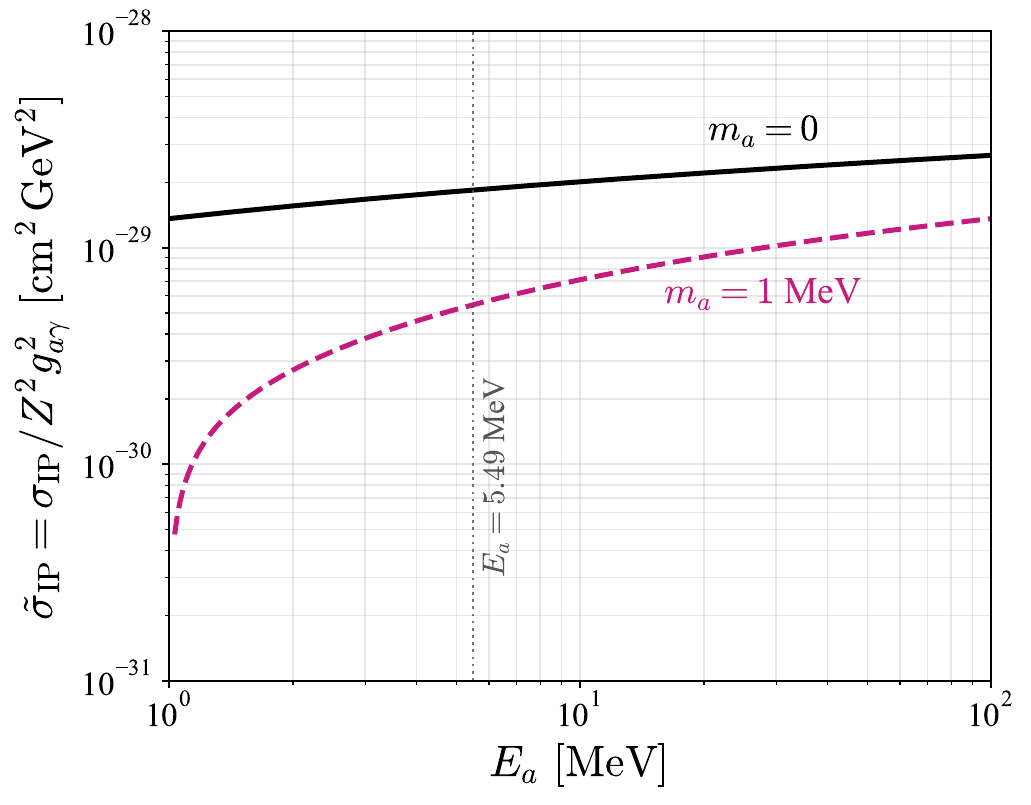}
\includegraphics[height=0.35\textwidth]{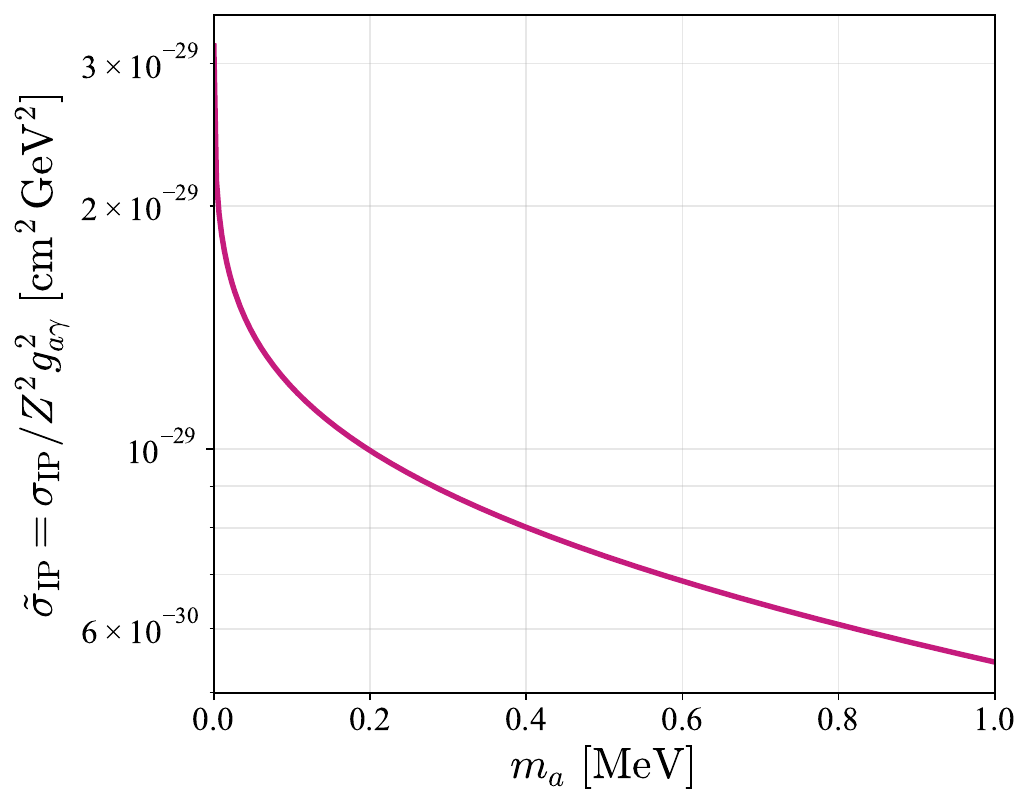}
\caption{
The reduced cross section $\tilde{\sigma}_{\rm IP} \equiv \sigma_{\rm IP}/Z^2g_{a\gamma}^2$ for the IP process $a + {}^A_Z X \to \gamma + {}^A_Z X$, as a function of the axion energy $E_a$ (left) and the axion mass $m_a$ with $E_a = 5.49$\,MeV (right). In the left panel, the solid black and dashed red lines are for $m_a = 0$ and 1\,MeV, respectively, and the fixed solar axion energy of $E_a = 5.49$\,MeV in this paper is indicated by the vertical dotted line.
}
\label{fig:primakoff}
\end{figure}

The axion-photon coupling $g_{a\gamma}$ can induce the IP
process $a + {}^A_Z X \to \gamma + {}^A_Z X$, in which a
solar axion converts into a photon in the Coulomb field of
a nucleus. 
The Feynman diagram is shown in \gfig{fig:primakoff_feynman}, 
and the corresponding total cross section is~\cite{BOREXINO:2025dbp, Avignone:1988bv} 
\begin{equation}
    \sigma_{\rm IP} \ = \ \frac{Z^2 \alpha g_{a\gamma}^2 }{2}
    \left[\frac{1 + \beta_a^2}{2\beta_a^2}
    \ln\!\left(\frac{1 + \beta_a}{1 - \beta_a}\right)
    - \frac{1}{\beta_a}\right] \,.
    \label{eq:sigma_pc}
\end{equation}
Similarly to the $e^+ e^-$ PP induced by the coupling $g_{ae}$, the cross section for the IP process scales with $Z^2$. As a result, the xenon-based detectors enjoy the enhancement factors of $(54/6)^2 = 81$ and $(54/8)^2 \simeq 45.6$ with respect to carbon and oxygen, respectively. For illustration, the reduced cross section $\tilde{\sigma}_{\rm IP} \equiv \sigma_{\rm IP}/Z^2 g_{a\gamma}^2$ is shown in the left panel of \gfig{fig:primakoff} as a function of the axion energy $E_a$, where the solid black and dashed red lines are for the values of $m_a = 0$ and 1\,MeV, respectively. The cross section $\tilde\sigma_{\rm IP}$ as a function of $m_a$ for the fixed energy of $E_a = 5.49$\,MeV is presented in the right panel of \gfig{fig:primakoff}. As implied by this figure, the cross section $\sigma_{\rm IP}$ increases with $E_a$ and decreases with $m_a$.

\subsection{Sensitivities}
\label{sec:sens_gay}

\begin{figure}[t]
\centering
\includegraphics[width=0.7\textwidth]{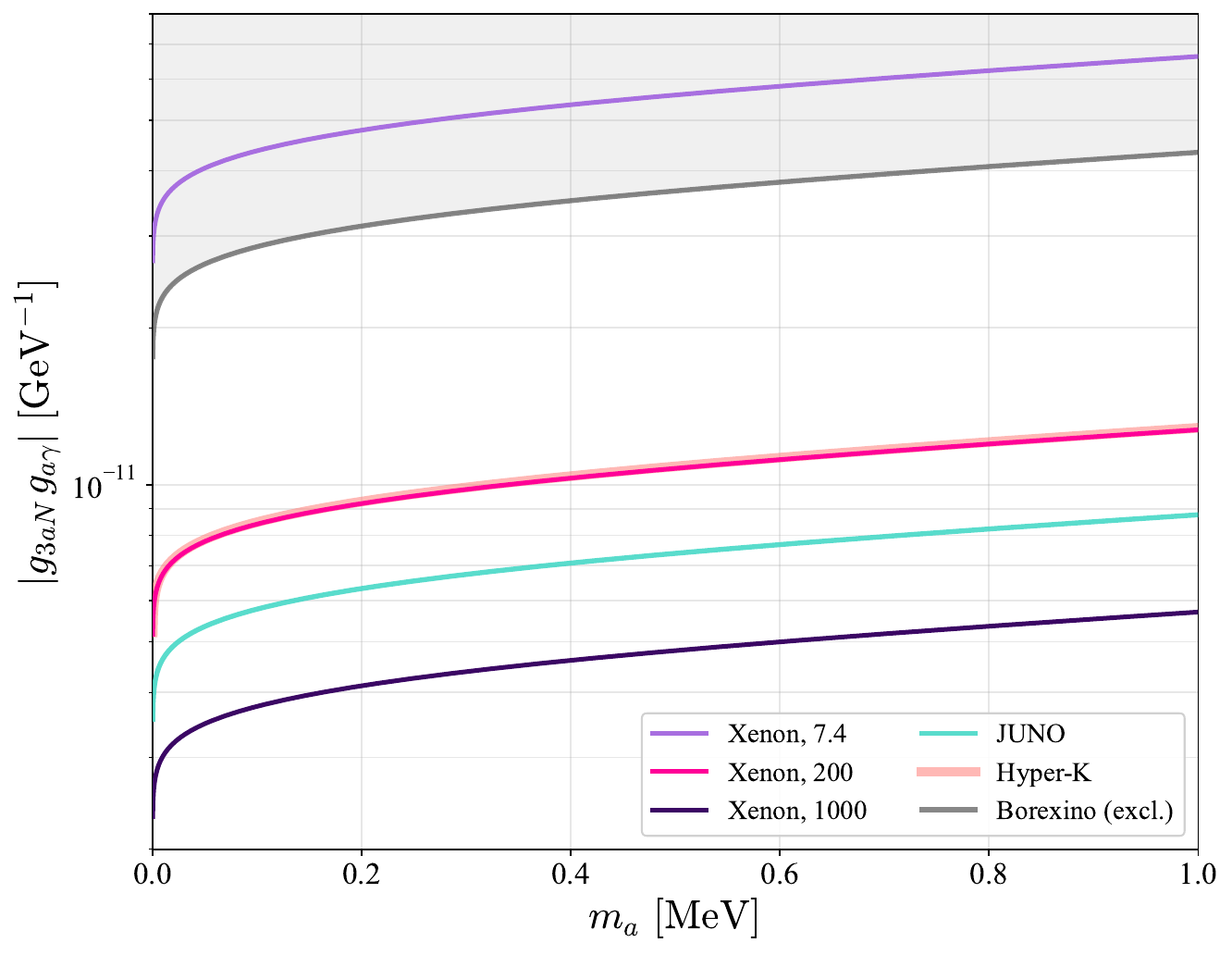}
\caption{
The 90\%\,C.L. sensitivities of $|g_{3aN} g_{a\gamma}|$ at the xenon experiments with the exposures of $7.4\,\mathrm{ton}\cdot\mathrm{yr}$ (light purple), $200\,\mathrm{ton}\cdot\mathrm{yr}$ (magenta) and $1000\,\mathrm{ton}\cdot\mathrm{yr}$ (dark purple) as functions of the axion mass $m_a$. Also shown are the current Borexino limit (shaded gray)~\cite{BOREXINO:2025dbp}, and the projections for JUNO ($1.62\times10^5\,\mathrm{ton}\cdot\mathrm{yr}$, light green) and Hyper-K ($1.87\times10^6\,\mathrm{ton}\cdot\mathrm{yr}$, light orange). The main results are collected in Table\,\ref{tab:comparison_gagamma}. 
}
\label{fig:sens_gay}
\end{figure}

The sensitivity estimation follows again the event rate formula in \geqn{eq:signal}. 
For the xenon nuclei $N_T = N_{\mathrm{Xe}}$, the expected signal rates scale with $g_{a\gamma}^2 g_{3aN}^2$, and the 90\%\,C.L. sensitivity is,
\begin{equation}
    |g_{3aN} g_{a\gamma}| = \sqrt{\frac{S_\mathrm{lim}}{K_{\rm IP}}}\,.
    \label{eq:sensitivity_gay}
\end{equation}
The 90\%\,C.L. sensitivity of $|g_{3aN} g_{a\gamma}|$ at xenon experiments with the exposures of  7.4, 200 and 1000\,ton$\cdot$yr are shown in \gfig{fig:sens_gay} as functions of axion mass $m_a$ in light purple, magenta and dark purple, respectively. 
In the limit of vanishing axion mass ($m_a \simeq 0$), the sensitivities can reach $|g_{3aN} g_{a\gamma}| \simeq 3.51 \times 10^{-11}\,\mathrm{GeV}^{-1}$, $6.76 \times 10^{-12}\,\mathrm{GeV}^{-1}$ and $3.02 \times 10^{-12}\,\mathrm{GeV}^{-1}$, respectively.\footnote{As the cross section $\sigma_{\rm IP}$ in Eq.~(\ref{eq:sigma_pc}) diverges in the limit of $m_a \simeq 0$, the sensitivities for $m_a \simeq 0$ in the IP process are actually evaluated at $m_a = 10$\,keV for reference in this paper.} 
For simplicity we make the na\"{i}ve zero-background assumption again to set the limit $S_{\rm lim} = 2.3$.

The current limit from Borexino is presented as the gray shaded region in \gfig{fig:sens_gay}~\cite{BOREXINO:2025dbp} (see also Ref.\,\cite{Borexino:2012guz}). With an exposure of 1587\,ton$\cdot$yr, the coupling $|g_{3aN} g_{a\gamma}|$ is constrained up to $2.3 \times 10^{-11}\,\mathrm{GeV}^{-1}$, precluding the xenon sensitivity with the exposure of 7.4\,ton$\cdot$yr.  
The axion-induced IP process produces a monochromatic 5.49\,MeV photon, whose energy deposition in the neutrino detectors such as JUNO and Hyper-K is expected to be similar to the $e^+e^-$ pairs induced by the coupling $g_{ae}$. 
Since the detector response in the MeV range is primarily sensitive to the total deposited energy rather than to the type of the incident particle, the background around 5.49\,MeV, and hence the upper limit $S_\mathrm{lim}$ on the number of signal events, are expected to be the same as in \gsec{sec:sens_gae}. 
\begin{table}[h]
\centering
\caption{
The 90\%\,C.L. sensitivities of xenon experiments to the 5.49\,MeV solar axions through the couplings $|g_{3aN}g_{a\gamma}|$, with axion mass $m_a \simeq 0$ or $m_a = 1$\,MeV. Also shown are the current limits from Borexino~\cite{BOREXINO:2025dbp} and the prospects at JUNO and Hyper-K. See text and Fig.\,\ref{fig:sens_gay} for more details. 
} \vspace{5pt}
\label{tab:comparison_gagamma}
\begin{tabular}{cccccc}
\hline
\multirow{2}{*}{Experiment} & \multirow{2}{*}{Target} & Exposure & Efficiency & $|g_{3aN}g_{a\gamma}|$ [GeV$^{-1}$] & $|g_{3aN}g_{a\gamma}|$ [GeV$^{-1}$] \\
& & (ton$\cdot$yr) & (\%) & at $m_a\simeq0$ & at $m_a=1$\,MeV \\
\hline
Borexino & $\mathrm{C_9H_{12}}$ & 1587 & 17.5 & $2.30\times10^{-11}$~\cite{BOREXINO:2025dbp} & $4.34\times10^{-11}$ \\
JUNO & $\mathrm{C_{19}H_{32}}$ & $1.62\times10^5$ & $\sim100$ & $4.64\times10^{-12}$ & $8.76\times10^{-12}$ \\
Hyper-K & $\mathrm{H_2O}$ & $1.87\times10^6$ & $\sim100$ & $6.82\times10^{-12}$ & $1.29\times10^{-11}$ \\
\hline
\multirow{3}{*}{\makecell{ xenon \\ experiments } } & \multirow{3}{*}{Xe} & 7.4 & \multirow{3}{*}{$\sim100$} & $3.51\times10^{-11}$ & $6.63\times10^{-11}$ \\
 & & 200 & & $6.76\times10^{-12}$ & $1.28\times10^{-11}$ \\
 & & 1000 & & $3.02\times10^{-12}$ & $5.70\times10^{-12}$ \\
\hline
\end{tabular}
\end{table}
With the exposures of $1.62\times10^{5}$ and $1.87\times10^{6}\,\mathrm{ton}\cdot\mathrm{yr}$ and the detection efficiency of $\sim100\%$, the reaches of $|g_{3aN} g_{a\gamma}|$ at $m_a\simeq 0$ are $4.64\times10^{-12}$\,GeV$^{-1}$ for JUNO and $6.82\times10^{-12}\,\mathrm{GeV}^{-1}$ at Hyper-K, respectively, as indicated by the light green and light orange lines in \gfig{fig:sens_gay}. For convenience, the current constraints on $|g_{3aN} g_{a\gamma}|$ from Borexino as well as the future prospects at the xenon experiments, JUNO and Hyper-K at $m_a \simeq 0$ and $m_a = 1$\,MeV are all collected in \gtab{tab:comparison_gagamma}. Although the exposure at xenon with 200\,ton$\cdot$yr is orders of magnitude smaller than that at JUNO and Hyper-K, benefiting from the $Z^2$ enhancement, 
the xenon experiments can achieve sensitivities on $|g_{3aN} g_{a\gamma}|$ comparable to those of the large neutrino detectors: the reach at $200\,\mathrm{ton}\cdot\mathrm{yr}$ is weaker than that of JUNO by a factor of $1.5$, while essentially matching that of Hyper-K. With a larger exposure of $1000\,\mathrm{ton}\cdot\mathrm{yr}$, the xenon sensitivities become better than those of JUNO and Hyper-K by a factor of $1.5$ and $2.3$, respectively. 

For completeness, we collect here some relevant constraints on the axion couplings to nucleons and photons, although they cannot be directly compared with our results in this paper. 
\begin{itemize}
    \item The 14.4\,keV solar axions from the M1 transition of ${}^{57}\mathrm{Fe}$ can induce photon signals at the DM direct detection experiments via the IP effect.
    Such measurement can therefore set limits on $|g_{aN}^{\rm eff} g_{a\gamma}|$
    with $g_{aN}^{\rm eff}$ again being a model specific coupling. The bound of $2.44\times10^{-16}\,\mathrm{GeV}^{-1}$ has been obtained for a massless axion by reanalyzing the public XENONnT data~\cite{Geng:2025yoj}.
    
    \item The solar axions can induce resonant absorption of the M1 transition of $^{169}$Tm at 8.41\,keV, and the constraint of $|g_{a\gamma}(g_{0aN}+g_{3aN})| \le 1.44\times10^{-14}\,\mathrm{GeV}^{-1}$ is obtained on the axion couplings at the 90\%\,C.L. \cite{Abdelhameed:2020hys}.

\end{itemize}

The coupling $g_{a\gamma}$ is also tightly constrained by 
the CMB and BBN observations~\cite{Masso:1995tw,Millea:2015qra,Depta:2020wmr,Depta:2020zbh,Langhoff:2022bij,Balazs:2022tjl,Cheng:2025cmb,Yin:2025amn}, the X-rays, $\gamma$-ray and extragalactic background light searches~\cite{Cadamuro:2011fd,Thorpe-Morgan:2020rwc,Calore:2022pks,Porras-Bedmar:2024uql}, 
supernovae~\cite{Masso:1995tw,Payez:2014xsa,Millea:2015qra,Jaeckel:2017tud,Lucente:2020whw,Calore:2020tjw,Caputo:2021rux,Balazs:2022tjl,Hoof:2022xbe,Ferreira:2022xlw,Diamond:2023scc,Muller:2023vjm,Muller:2023pip,Candon:2025ypl,Caputo:2022mah,Fiorillo:2025yzf}, neutron star mergers~\cite{Diamond:2023cto,Dev:2023hax} and the beam-dump experiments~\cite{Dobrich:2015jyk,Capozzi:2023ffu}.

\section{Conclusion}
\label{sec:conclusion}

In this paper, we have estimated the sensitivity of ongoing and future xenon detectors, such as XENONnT, PandaX-xT, LZ and XLZD, 
to the 5.49\,MeV solar axions 
for axion mass $m_a \in [0, 1]\,\mathrm{MeV}$. Such
monochromatic axions can be produced from the $p + d \rightarrow {}^3$He process by the isovector axion-nucleon coupling $g_{3aN}$. We also recast some of the current Borexino limits on the axion couplings and revisit the prospects at the neutrino experiments JUNO and Hyper-K. 
The main results of this paper can be summarized as follows.
\begin{itemize}
    \item In the presence of the axion-electron coupling $g_{ae}$, there are three detection channels, i.e. the axion-induced $e^+ e^-$ PP, IC and the AE process. The cross section for the PP channel is largely enhanced when the axion mass is large for all the targets involved in this paper, up to a factor of $\sim 100$ at $m_a=1$\,MeV with respect to the massless case. Furthermore, the PP channel at xenon is enhanced by the atomic number $Z^2$, with respect to the lighter targets of carbon in Borexino and JUNO as well as oxygen in Hyper-K. Given these two enhancement factors, the signals at xenon experiments are dominated by the PP process, with subleading contributions from the IC and AE channels.     
    This makes the next-generation xenon experiments as competitive as the large neutrino detectors, although the exposures of xenon experiments are two or three orders of magnitude smaller. 

    The estimated sensitivities of $|g_{3aN}g_{ae}|$ at the xenon experiments with the three benchmark values of exposures of 7.4, 200 and 1000 ton$\cdot$yr, the recast Borexino limits and the prospects at JUNO and Hyper-K are presented in \gfig{fig:sens_gae} and \gtab{tab:comparison_gae}. 
    With an exposure of 7.4 ton$\cdot$yr, the xenon experiments can probe the couplings $|g_{3aN}g_{ae}|$ down to $8.28 \times10^{-14}$, which has been precluded by the recast Borexino limit $2.91\times10^{-14}$. With larger exposures of 200 and 1000\,ton$\cdot$yr, the sensitivities can be improved up to $1.59 \times10^{-14}$ and $7.12 \times10^{-15}$, respectively, which are comparable to or even better than the expected projections of $1.00 \times10^{-14}$ at JUNO and $1.49 \times10^{-14}$ at Hyper-K.

\item Given the axion-photon coupling $g_{a\gamma}$, the signal emerges as the IP process and the corresponding cross section also scales as $Z^2$. The recast Borexino limit and the prospects at the xenon experiments, JUNO and Hyper-K are shown in \gfig{fig:sens_gay} and \gtab{tab:comparison_gagamma}. The xenon sensitivity of $|g_{3aN}g_{a\gamma}|$ can reach $3.51\times10^{-11}\,\mathrm{GeV}^{-1}$ with the exposures of 7.4\,ton$\cdot$yr, which is precluded by the current Borexino limit $2.3\times10^{-11}\,\mathrm{GeV}^{-1}$. With the larger exposures of 200 and 1000\,ton$\cdot$yr, the sensitivities can be improved up to $6.76\times10^{-12}\,\mathrm{GeV}^{-1}$ and $3.02\times10^{-12}\,\mathrm{GeV}^{-1}$, respectively, which are comparable to or even better than the projections of $4.64\times10^{-12}\,\mathrm{GeV}^{-1}$ at JUNO and $6.82\times10^{-12}\,\mathrm{GeV}^{-1}$ at Hyper-K.
 
\end{itemize}

In short, while the future large-scale experiments such as JUNO and Hyper-K can reach unprecedented sensitivities as a result of their enormous exposures, the next-generation xenon detectors can provide a complementary and competitive probe of the MeV scale solar axions with a far smaller target mass.

\section*{Acknowledgements}

The authors would like to thank Ke Han and Jianglai Liu for useful discussions. The work of RF and YZ is supported by the National Natural Science Foundation of China (NSFC) under Grant No. 12175039. The work of SFG and OT is supported by the National NSFC under the Grant No. of 12375101 and 12425506. This work is also supported by
State Key Laboratory of Dark Matter Physics and the Fundamental Research Funds for the Central Universities. 

\bibliographystyle{utphysGe}
\bibliography{ref}

\end{document}